\documentclass[aps,prl,showpacs,floatfix,superscriptaddress,twocolumn]{revtex4-1}
\usepackage{xr}
\usepackage{amsmath,amssymb,physics}
\usepackage[toc,page]{appendix}
\usepackage{verbatim,graphicx,mathtools}
\usepackage{txfonts}
\usepackage{color}
\usepackage[all]{xy}
\usepackage[hidelinks]{hyperref}
\usepackage{subfigure,float}

\hypersetup{
  colorlinks   = true, 
  urlcolor     = blue, 
  linkcolor    = blue, 
  citecolor   = blue 
}

\newcommand{\be}{\begin{eqnarray}}
\newcommand{\ee}{\end{eqnarray}}

\newcommand{\bmat}{\left ( \begin{array}{cc} }
	\newcommand{\emat}{\end{array} \right ) }

\def\Tr{\textrm{Tr}}

\newcommand{\half}{{1\over2}}
\newcommand{\sgn}{{\rm sgn}}

\newcommand{\beq}{\begin{equation}}
\newcommand{\beqs}{\begin{equation*}}
\newcommand{\eeq}{\end{equation}}
\newcommand{\eeqs}{\end{equation*}}

\allowdisplaybreaks

\begin{document}

\title{Chaotic-Integrable Transition in the Sachdev-Ye-Kitaev Model}
\author{Antonio M. Garc\'\i a-Garc\'\i a}
\affiliation{Shanghai Center for Complex Physics, Department of Physics and Astronomy, Shanghai Jiao Tong University, Shanghai 200240, China}
\email{amgg@sjtu.edu.cn}
\author{Bruno Loureiro}
\affiliation{TCM Group, Cavendish Laboratory, University of Cambridge, JJ Thomson Avenue, Cambridge, CB3 0HE, UK}
\email{bl360@cam.ac.uk}
\author{Aurelio Romero-Berm\'udez}
\affiliation{Instituut-Lorentz for Theoretical Physics, $\Delta ITP$, Leiden University, Niels Bohrweg 2, Leiden 2333CA, The Netherlands}
\email{romero@lorentz.leidenuniv.nl}
\author{Masaki Tezuka}
\affiliation{Department of Physics, Kyoto University, Kyoto 606-8502, Japan}
\email{tezuka@scphys.kyoto-u.ac.jp}

\begin{abstract} 
Quantum chaos is one of the distinctive features of the Sachdev-Ye-Kitaev (SYK) model, $N$ Majorana fermions in $0+1$ dimensions with infinite-range two-body interactions, which is attracting a lot of interest as a toy model for holography. Here we show analytically and numerically that a generalized SYK model with an additional one-body infinite-range random interaction, which is a relevant perturbation in the infrared, is still quantum chaotic and retains most of its holographic features for a fixed value of the perturbation and sufficiently high temperature. However a chaotic-integrable transition, characterized by the vanishing of the Lyapunov exponent and spectral correlations given by Poisson statistics, occurs at a temperature that depends on the strength of the perturbation. We speculate about the gravity dual of this transition.
\end{abstract}
\maketitle


Motivated by its potential applications in high-energy and condensed matter physics, and also because of its simplicity, research on fermionic models with infinite-range random interactions \cite{bohigas1971,bohigas1971a,french1970,french1971,mon1975,benet2003,kota2014,sachdev1993,sachdev2010}, now generally called Sachdev-Ye-Kitaev (SYK) models \cite{kitaev2015,jensen2016,maldacena2016,sachdev2015}, has flourished in recent times \cite{almheiri2015,maldacena2016a,engels2016,bagrets2016,cenke2017,danshita2016,jensen2016,jevicki2016,jian2017,magan2016,magan2016b,witten2016,you2016,garcia2016,cenke2017,garcia2017,klebanov2017,bagrets2017,cotler2016,altman2017,kanazawa2017,krishnan2017,krishnan2017a}.
 Interesting research lines currently being investigated include not only applications in holography \cite{kitaev2015,jensen2016,maldacena2016,sachdev2015} but also in random matrix theory \cite{you2016,garcia2016,cotler2016,garcia2017,kanazawa2017,krishnan2017a}, possible experimental realizations \cite{danshita2016,garcia-alvarez2016,Pikulin2017}, 
and extensions involving nonrandom couplings \cite{witten2016,klebanov2017}, higher spatial dimensions \cite{gu2016,jian2017,cenke2017,altman2017,davison2017}, and several flavors \cite{gross2017}. \\
A natural question to ask \cite{witten2016,altman2017,jian2017,cenke2017,gross2017,davison2017,gu2016} is to what extent holographic properties are present in generalized SYK models. For instance, similar features are observed for nonrandom couplings \cite{witten2016} and in higher-dimensional realizations of the SYK \cite{gu2016,davison2017} model. However, in some cases, the addition of more fermionic species can induce a transition to a Fermi liquid phase \cite{altman2017} or a metal-insulator transition \cite{jian2017,cenke2017}, which, at least superficially, spoils a holographic interpretation. 
Here we study the stability of chaos and holographic features of a generalized SYK model consisting of $N$ fermions in $0+1$ dimension with infinite-range two-body random interaction perturbed by a one-body random term
 \begin{equation}\label{hami}
\hspace{-1mm}H \, = \, \frac{1}{4!} \sum_{i,j,k,l=1}^N J_{ijkl} \, \chi_i \, \chi_j \, \chi_k \, \chi_l \, +\, \frac{i }{2!} \sum_{i,j=1}^N \kappa_{ij} \, \chi_i \, \chi_j \, ,
\end{equation}
where $\chi_i$ are Majorana fermions so
$
\{ \chi_i, \chi_j \} = \delta_{ij}$.
The couplings $J_{ijkl}$ and $\kappa_{ij}$ are Gaussian-distributed random variables
with zero average and standard deviation $\frac{\sqrt{6}J}{N^{3/2}}$, $\frac{\kappa}{\sqrt{N}}$ respectively
\cite{kitaev2015,maldacena2016}. 
We study the model \cite{maldacena2016,sachdev2015} by introducing replica fields, averaging over disorder and decoupling the replica fields by two Hubbard-Stratonovich transformations that allow the integration of the original fermionic variables.
 The resulting partition function is expressed in terms of the bilocal fields $G(\tau_1,\tau_2)$ and $\Sigma(\tau_1,\tau_2)$: $Z=\int [DG][D\Sigma]e^{-N S_\mathrm{eff}}$, where 
 \\
\begin{align}
S_\mathrm{eff}&=-\frac{1}{2}\Tr \log( \partial_\tau-\Sigma)+{1\over2}\int \dd \tau\dd\tau'\left[G(\tau,\tau')\Sigma(\tau,\tau')\right. \nonumber\\
&\left.-{J^2\over 4}G(\tau,\tau')^4-{\kappa^2\over 2}G(\tau,\tau')^2\right]\,.\label{seff}
\end{align}
The saddle point equations in imaginary time, which become exact in the large-$N$ limit of interest, are: 
\begin{equation}\label{aa3}
G_n^{-1}={-i\omega_n -\Sigma_n }\,,\  \Sigma(\tau)=-J^2 G(\tau)^2G(-\tau)+\kappa^2G(\tau)\,,
\end{equation}
where $\omega_n=(2\pi/\beta)(n+1/2)$,  $G_n\equiv G(i\omega_n)$ and $\Sigma_n\equiv \Sigma(i\omega_n)$.  
In the long time, strong coupling limit, where conformal symmetry holds, the solution of the Schwinger-Dyson (SD) equations is dominated by the one-body term and therefore \cite{kitaev2015,maldacena2016} the zero-temperature entropy always vanishes. The low-temperature limit of the specific heat is directly related to the leading correction to the conformal Green's function. A simple power-counting argument in the SD equations suggests that it has contributions from both terms. Therefore we expect the specific heat still to be linear  
with a slope $c$ that may depend on $\kappa$. We confirm these results by exact diagonalization of Eq.~(\ref{hami}) with $J = 1$. 
For a given set of parameters, we have obtained at least $10^6$ eigenvalues.  We have computed, following \cite{garcia2016,cotler2016}, the entropy at zero-temperature $s_0$ and the specific heat by using standard thermodynamic relations and a finite-size scaling analysis.
As was expected, we have found a vanishing $s_0$ for any $\kappa$ and a linear specific heat with $c \propto N f(\kappa)$ with $f \sim 0.5/\kappa$ for small $\kappa$ and a steady increase for larger $\kappa$.
We note that all these features, including $s_0 = 0$ \cite{gouteraux2011}, are consistent with the existence of a gravity dual.  
We have confirmed these results by an explicit evaluation of the free energy from Eqs. (\ref{seff}) and (\ref{aa3}). See Suppermental Material, Secs. A and B for more details. 
Next we employ level statistics \cite{guhr1998} to investigate the effect of the one-body perturbation in the quantum chaotic features of the model.\\

\begin{figure}
	\centering

\includegraphics[scale=0.42]{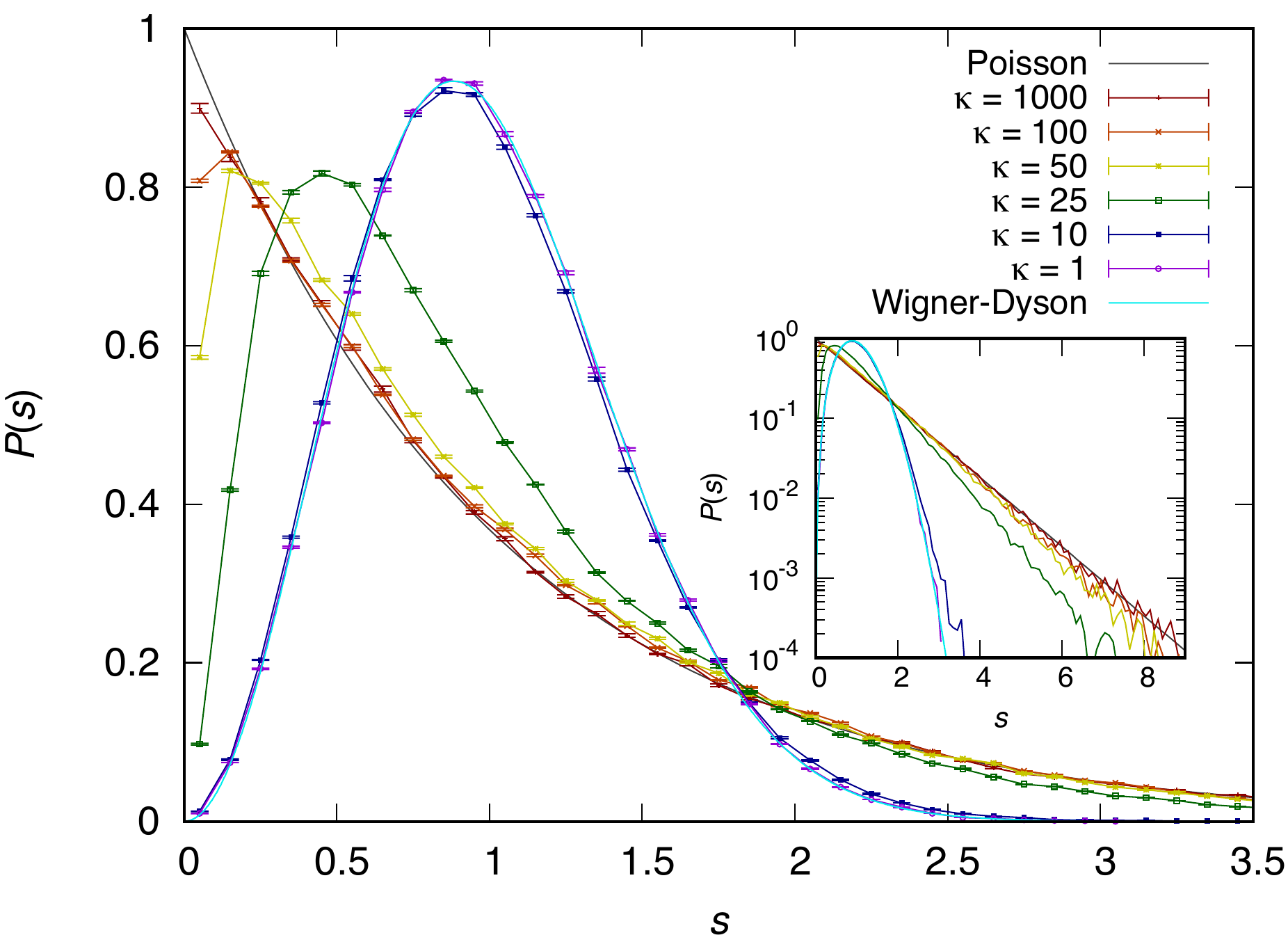}
	
	\vspace{-2mm}
	
	\includegraphics{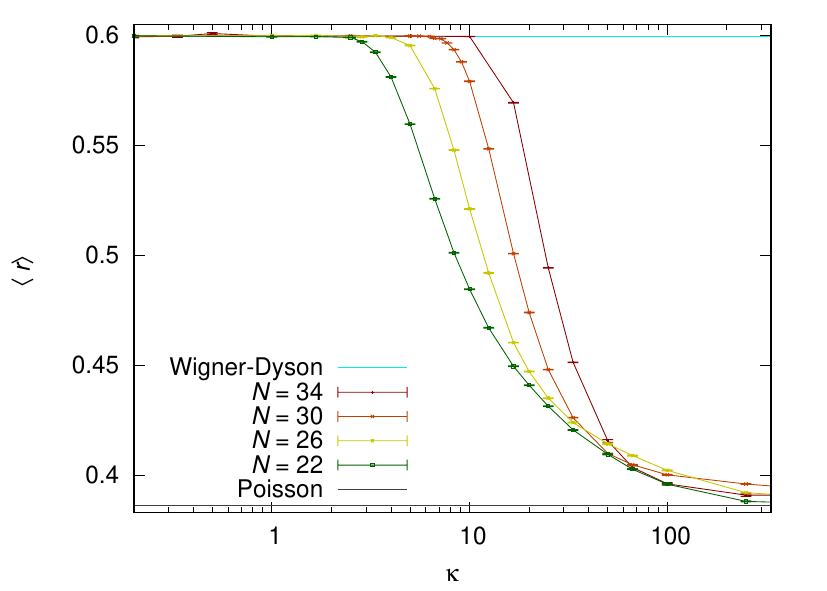}
		
		\vspace{-4mm}
	\caption{Upper: $P(s)$ for $N = 34$ and different $\kappa$'s with $\kappa$ in units of $J=1$. We clearly observe a crossover from Wigner-Dyson (WD) to Poisson statistics as $\kappa$ increases. Lower: Finite-size scaling analysis of the averaged adjacent gap ratio $\langle r \rangle$ Eq.~(\ref{eq:agr}) as a function of $\kappa$ for different $N$'s. 
	 For sufficiently large $N$ we observe a crossing at $\kappa_c \approx 66$ which suggests the existence of a chaotic-integrable transition. Results for larger $N$ would be necessary to confirm it. See main text for an explanation of the absence of crossing for small $N$. Both $P(s)$ and $\langle r \rangle$ were computed by using ensemble and spectral average in a window comprising $10\%$ of the eigenvalues around the center of the spectrum. Results are robust to changes in the percentage of eigenvalues provided that the spectrum edges are avoided.} 
	\label{fig1}
\end{figure}

\begin{figure}
	\centering
	\resizebox{0.45\textwidth}{!}{\includegraphics{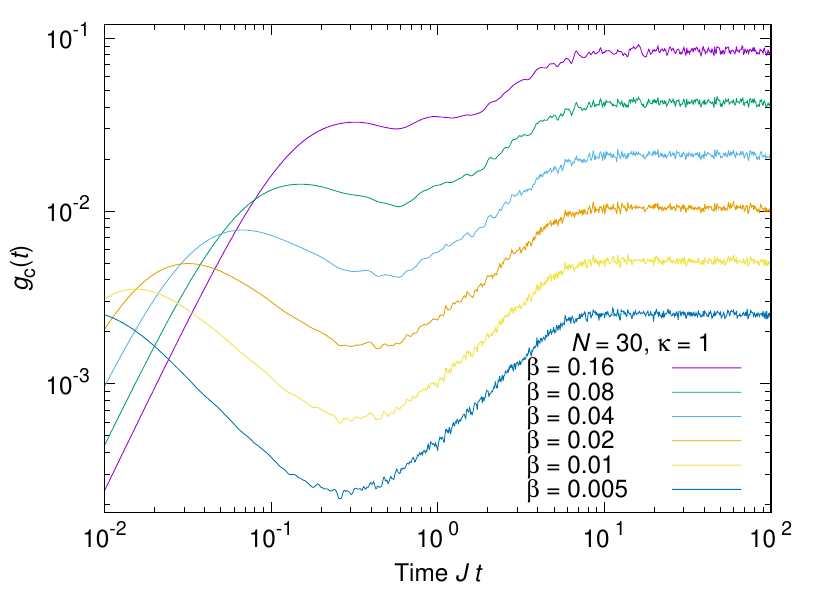}} 
\resizebox{0.45\textwidth}{!}{\includegraphics{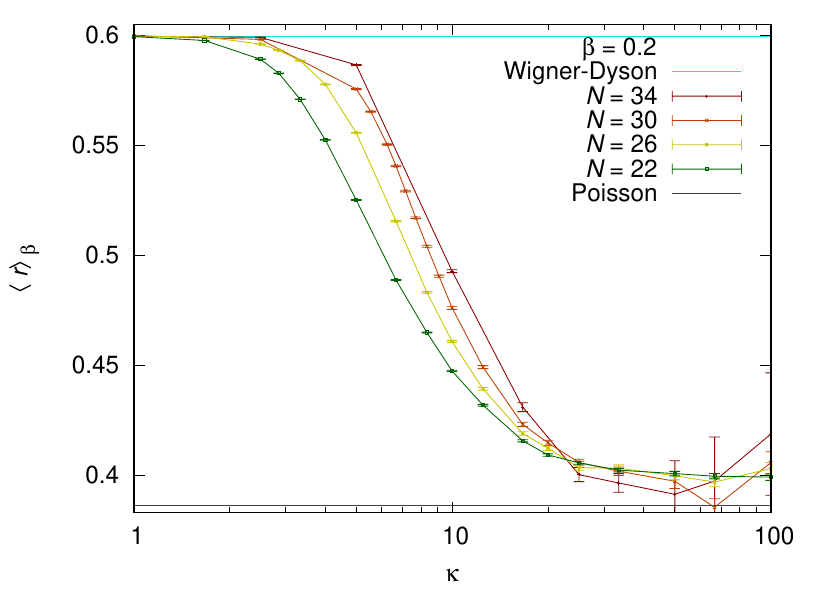}}
		\vspace{-4mm}
	\caption{
		Upper: Connected spectral form factor $g_c(t)$ for the unfolded spectrum from Eq.~(\ref{eq:gt}) for $N = 30$, $J = \kappa = 1$ and different $\beta$'s. For small $\beta$, we observe the correlation hole \cite{alhassid1992,Torres-Herrera2017} followed by a ramp typical of quantum chaotic systems. As $\beta$ increases, that probes the tail of the spectrum, results are not conclusive. Lower: A finite-size scaling analysis, also with $J=1$, of the average adjacent gap ratio $\langle r \rangle_\beta$ Eq. \eqref{eq:agr}, where, unlike the previous figure, the average is weighted by the function
$\exp(-\beta (E_i + 2E_{i+1} + E_{i+2})/4)$ but the spectrum is not unfolded. Here we have excluded the ten smallest eigenvalues from the analysis because $\langle r \rangle$ for such eigenvalues at the spectral tail is anomalous high even in the large $\kappa$ limit. For details, see the supplemental material. For $\beta = 0.2$, which probes the low energy part of the spectrum, we observe a crossing at $\kappa_c \approx 25$ that seems to indicate a chaotic-integrable transition. However the size dependence is too weak to confirm the existence of the transition.} 	\label{fig2} 
\end{figure}
{\it Level statistics.--}
For that purpose, we compute, from the exact diagonalization of Eq.~(\ref{hami}), the level spacing distribution $P(s)$,  the probability to find two consecutive eigenvalues $E_{i}, E_{i+1}$ at a distance $s = (E_{i+1}-E_{i})/\Delta$, that probes the system dynamics for times of the order of the Heisenberg time $\sim \hbar/\Delta$ with $\Delta$ the mean level spacing.    
For an insulator, or a generic integrable system, it is given by Poisson statistics, 
$P_\mathrm{P}(s) = e^{-s}$ \cite{guhr1998},  
while for a quantum chaotic system it is given by WD statistics \cite{mehta2004} which is well approximated by the Wigner surmise, 
$
P_\mathrm{W}(s) \approx \frac{32}{\pi^2}s^{2}\exp(-4 s^2/\pi),
$
for systems with broken time reversal invariance \cite{guhr1998}.  For a meaningful comparison with these predictions, unfolding the spectrum \cite{guhr1998} is necessary so that $\Delta = 1$ by a local fitting of the numerical spectral density by a smooth function which is subsequently employed to rescale the spectrum.

Results for $P(s)$ as a function of $\kappa$, depicted in Fig.~\ref{fig1}, show a gradual crossover from Poisson to WD statistics as $\kappa$ decreases, which is also observed in the tail of the distribution (see inset). Unlike the standard SYK model \cite{you2016,garcia2016}, the one-body term breaks time reversal invariance for all $N$.  
As was mentioned previously, for $J = 0$ the Hamiltonian is effectively noninteracting which suggests that Poisson statistics applies in the $N \to \infty$ limit.  
In order to determine whether Poisson statistics is robust for $J \gg \kappa$, and therefore a transition occurs at a finite $\kappa$, we carry out a finite-size scaling analysis employing as scaling variable the adjacent gap ratio \cite{luitz2015,oganesyan2007,bertrand2016}, 
\begin{equation}
r_i = \frac{\min(\delta_i, \delta_{i+1})}{\max(\delta_i, \delta_{i+1})} 
\label{eq:agr}
\end{equation}
for an ordered spectrum $E_{i-1} < E_i < E_{i+1}$  where $\delta_i = E_i - E_{i-1}$.
The average adjacent gap ratio for a Poisson distribution is $\left\langle r \right\rangle_\mathrm{P} = 2\ln(2) - 1 \approx 0.386$ while for WD statistics 
is $\approx 0.599$ \cite{atas2016}.  
In Fig.~\ref{fig1}, $\langle r \rangle$ is depicted as a function of $\kappa$ for different $N$'s. Only $10\%$ of the total number of eigenvalues, located around the center of the spectrum, are employed in the calculation. We stick to values of $N$ for which time reversal symmetry is broken even in the absence of one-body term in Eq.~(\ref{hami}). 
As was expected, except for $\kappa \gg 10$, $\langle r \rangle$ is very close to the WD result for any $N$.  
Only for $\kappa \geq 25$, a crossover to Poisson statistics is observed. For $N \leq 22$, we did not observe a crossing point in the plot, and moreover, $\langle r \rangle$ gradually approaches the WD prediction, which suggests that the system is quantum chaotic for any $\kappa$. However, for $N \geq 30$, we observe a crossing at $\kappa = \kappa_c \approx 60$. This is an indication of a transition from chaos to integrability, though it would be necessary to explore larger $N$'s to confirm it. 
A possible explanation for the different behavior for small $N$ is that the lowest eigenvalues (the most infrared part of the spectrum) are strongly correlated. The reason is that the one-particle sector for $\kappa \to \infty$, which controls the lowest energy properties, is known \cite{cotler2016} to be described by a skew-orthogonal random matrix. The number of eigenvalues related to one-particle states decreases exponentially with $N$, which would explain why its contribution is only relevant for sufficiently small $N$. 

We note the existence of the gravity dual is related to the properties of the model in the low-temperature, strong coupling limit described by the tail, not the bulk of the spectrum studied above. 
Moreover, we are also interested in the nature of level statistics for shorter timescales where random matrix theory predicts level rigidity \cite{guhr1998}.  
In order to investigate these issues, we compute the connected spectral form factor of the unfolded spectrum
\begin{equation}\label{eq:gt}
g_\mathrm{c}(t) \equiv \left \langle \frac{Z(t,\beta)Z^*(t,\beta)}{Z(0,\beta)^2}  \right \rangle - \left \vert \left \langle \frac{Z(t,\beta)}{Z(0,\beta}\right \rangle\right \vert^2
\end{equation}
where $Z(t,\beta) = \Tr e^{- \beta H - i H t}$ and $\beta > 0$. 
For quantum chaotic systems, and also for the unperturbed SYK model \cite{cotler2016,garcia2016}, we expect a correlation hole \cite{alhassid1992,kudrolli1994,cotler2016,Torres-Herrera2017,alt1997} for intermediate times followed by a ramp, related to the level rigidity observed in quantum chaotic systems \cite{guhr1998}. We note that, by increasing $\beta$, we probe the tail of the spectrum. Results, depicted in Fig.~\ref{fig2}, show that for $\kappa = 1$ the ramp is still observed for sufficiently small $\beta$. 
In order to clarify the situation for larger $\beta$, which probes the spectral correlations of the smallest eigenvalues, we again carry out a finite-size scaling analysis of the averaged adjacent gap ratio $\langle r \rangle$ [Eq. \eqref{eq:agr}], but the average is weighted by $\beta$ (see caption of Fig.~\ref{fig2}) so that the low energy part of the spectrum is singled out.

 A crossing seems to be observed (see lower plot of Fig.~\ref{fig2}) at $\kappa = \kappa_c \approx 25$. This is a signature of a chaos-integrable transition in the tail of the spectrum. However, results are not conclusive because the size dependence is weak for large $\kappa$. We only note that, in qualitative agreement with the results of next section, $\kappa_c$ for $\beta = 0.2$ is smaller than in Fig.~\ref{fig1}, where effectively $\beta \approx 0$. 
 It is worth mentioning that similar chaotic-integrable transitions in level statistics have been previously studied in the context of nuclear physics \cite{aberg1990} and quantum chaos \cite{jacquod1997,berkovits1998,kota2011a,kota2014} in somehow related models such as complex fermions with infinite-range interactions and a random diagonal one-body term or interacting systems with short-range interactions \cite{stasio1995}.

In summary, the finite-size scaling analysis is not fully conclusive to detect the chaotic-integrable transition.  
In order to confirm it, we investigate next out-of-time-order four-point correlation functions where quantum chaotic features are characterized by a finite Lyapunov exponent \cite{sekino2008,maldacena2015,larkin1969}.\\
{\it Out-of-time-order four-point correlation function.--}
In the semiclassical limit,     
the time evolution of certain out-of-time-order correlation function experiences a period of exponential growth \cite{larkin1969,berman1978} around the Ehrenfest time $t_* \sim \lambda_L^{-1}\log (\hbar/S_0)$, where $S_0$ is a typical action of the system and $\lambda_L$ is the classical Lyapunov exponent. 
By contrast, for nonchaotic systems, the growth of $t_*$ with $\hbar$ is only power law \cite{lai1993}. The application of these ideas in high-energy physics, where $\hbar$ is traded by a parameter $\sim 1/N$ that controls small quantum gravity corrections, has led to the proposal that black holes are quantum chaotic \cite{sekino2008} with a Lyapunov exponent that saturates a recently proposed universal bound $\lambda_L \leq 2\pi k_B T /\hbar$ \cite{maldacena2015}. We now study whether these chaotic features are present in Eq. \eqref{hami}.
We compute $\lambda_L$ from the following \cite{maldacena2015,polchinski2016,maldacena2016} out-of-time-order correlator
\begin{align}
\hspace{-2mm}F(t_1,t_2){\equiv}& {1\over N^2}\sum_{i,j}^N {\rm Tr}\left[\rho(\beta)^{1\over4}\chi_i(t_1)\rho(\beta)^{1\over4}\chi_j(0)\rho(\beta)^{1\over4}\chi_i(t_2)\rho(\beta)^{1\over4}\chi_j(0)\right]\nonumber\\
&\simeq G_R(t_{1})G_R(t_{2})+{1\over N}{\cal F}(t_1,t_2)+{\cal O} \Big({1\over N^{2}}\Big)\,,\label{eq:OTO_def}
\end{align}
where $\rho^{1/4}(\beta) = \left(\frac{e^{-\beta H}}{Z}\right)^{1/4}$ is inserted \cite{maldacena2015} along the thermal cycle to regularize the otherwise divergent operator. It is possible to show that ${\cal F}(t_1,t_2)$ satisfies
\begin{align}\label{eq:otos}
  \phantom{K_R(t_1,t_2,t_3,t_4}
    \mathllap{{\cal F}(t_1,t_2)} &=\hspace{-1mm} \int\hspace{-1mm}\dd t_3\dd t_4 K_R(t_1,t_2,t_3,t_4) {\cal F}(t_3,t_4)\,,\\
    \mathllap{K_R(t_1,t_2,t_3,t_4)} &=G_{R}(t_{1})G_{R}(t_{2})\hspace{-1mm}\left[3 J^2G_{lr}^2(t_{3}-t_4)\hspace{-0.5mm}+\hspace{-0.5mm}\kappa^2 \right],\label{eq:otos_K}
\end{align}
where $G_R(\omega)=G(i\omega_n\to \omega+i 0^+)$ is the retarded Green's function in real frequency and $G_{lr}(t)$ is the Wightman function obtained from $G_{lr}(\omega)={2ie^{-{\beta\omega/ 2}}\over 1+e^{-\beta \omega}}\Im(G_R(\omega))$. 
In order to compute these two-point functions we follow the strategy employed in Refs. \cite{maldacena2016,altman2017}. We  analytically continue $i\omega_n\to \omega+i 0^+$ the saddle point equations  (\ref{aa3}) and solve them using the spectral representation of the retarded Green's function. 
\begin{figure}
	\hspace{-5mm}
	\resizebox{0.5\textwidth}{!}{\includegraphics{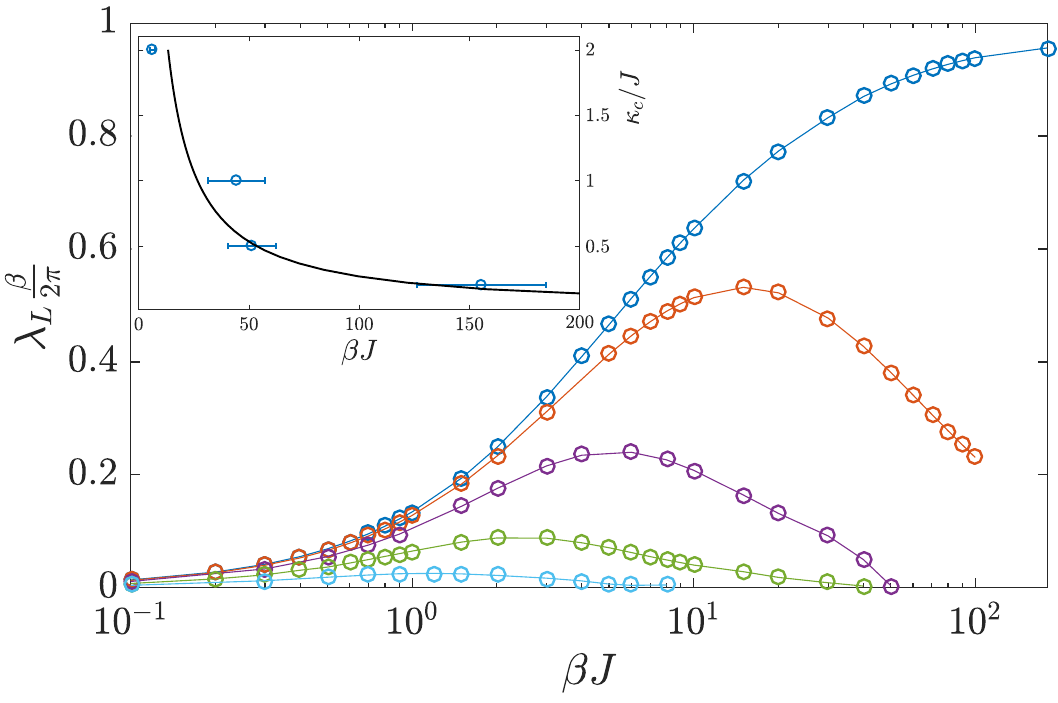}}
		\vspace{-3mm}
	\caption{Lyapunov exponent $\lambda_L$ for the model Eq. \eqref{hami} with $J = 1$ as a function of the inverse temperature $\beta = 1/T$ and $\kappa$. From top to bottom, $\kappa = 0,\ 0.2,\ 0.5,\ 1,\ 2$. A finite $\lambda_L$, which is a signature of quantum chaos, is observed for a fixed $\kappa$ and not too low temperature. For sufficiently low temperatures, and $\kappa > 0$, we identify a $T^*(\kappa)$ such that $\lambda_L = 0$ for any $T < T^*(\kappa)$ which signals a chaotic-integrable transition. (Inset) Critical value of $\kappa=\kappa_c$ at which the transition takes place as a function of $\beta$. Dots result from fitting the numerical data of the main plot near the transition. The solid line is the analytical expression from Eq.\eqref{eq:Lyapunov_kappa} valid in the large-$q$ limit. Agreement with the numerical data is reasonable for $\kappa/J\ll1$. }
	\label{fig:Lyapunov}%
\end{figure}
Substituting the ansatz  ${\cal F}(t_1,t_2)=e^{\lambda_L(t_1+t_2)/2}f(t_{12})$, where $t_{12}=t_1 -t_2$, into Eq.~(\ref{eq:otos}) and expressing it in the frequency domain, we obtain the following eigenvalue equation for $f(\omega)$,
\begin{equation}\label{eq:f_eigeneq}
\begin{split}
\hspace{-3mm}f(\omega'){=}\left|G^R\left(\omega'{+}i{\lambda_L\over 2}\right)\right|^2 \left[\kappa^2 f(\omega')
{+}
3J^2\hspace{-2mm}\int\hspace{-1mm} {\dd\omega\over 2\pi}
g_{lr}(\omega'{-}\omega)f(\omega)\right]\\
\end{split}
\end{equation}
where $\omega'=\omega_1-i\lambda_L/2$ and $g_{lr}(\omega)=\int \dd  te^{i\omega t}G_{lr}(t)^2$.
Finally, we compute $\lambda_L$ by imposing the existence of a nondegenerate eigenvalue equal to one so that Eq.~(\ref{eq:f_eigeneq}) is satisfied. 

Results, depicted in Fig.~\ref{fig:Lyapunov}, show the system displays chaotic behavior, namely, the Lyapunov exponent $\lambda_L$ is finite, for all studied values of $\kappa$ and sufficiently high temperature. However, even in the strong coupling limit, $\lambda_L$ never approaches the bound $\lambda_L = 2\pi k_B T /\hbar$. Indeed, for a given temperature, $\lambda_L$ decreases as $\kappa$ increases and eventually vanishes for sufficiently strong $\kappa$ or, for a fixed $\kappa$, for sufficiently low temperature. Therefore, quantum chaos is robust to the introduction of a relevant one-body perturbation but only if it is weak enough and the temperature is high enough. We now confirm these results analytically by studying the following model with $q/2$-body interactions,
\begin{align}\label{hami:largeq}
\hspace{-1mm}H = i^{q\over 2} \hspace{-4mm}\sum_{1\leq i_1<i_2<\dots<i_q\leq N}\hspace{-4mm} J_{i_1, i_2,\dots, i_q}\ \chi_{i_1}\chi_{i_2} \dots \chi_{i_{q}} +i\sum_{1\leq i<j\leq N} \kappa_{ij} \chi_i \chi_j \,,
\end{align}
\noindent where $\kappa_{ij}$ and $J_{i_1, i_2,\dots, i_q}$ are again Gaussian-distributed random variables with zero average and $q$-dependent variances $\frac{{\kappa}^2}{q N}$, $\frac{2^{q-1}}{q}\frac{(q-1)!{J}^2}{N^{q-1}}$, and $q \gg 1$, respectively. As before, we fix $J$ and use $\kappa$ and $\beta$ as the only parameters. The key insight is that the retarded kernel in this model, given by
\begin{align}
\label{eq:largeq:retarded_kernel}
\hspace{-4mm}K_{R}(t_1{,}t_2{,}t_3{,}t_4) {=} {G_{R}(t_1)G_{R}(t_2)\over q}\left[(q{-}1)J^2G_{lr}(t_3{-}t_4)^{q{-}2}{+}\kappa^2 \right]	
\end{align}
\noindent  simplifies considerably in the limit of $q\gg 1$, allowing for an analytical solution of the eigenvalue problem in Eq.\eqref{eq:otos}. First, we proceed in the same way as for the model \eqref{hami} to find an effective action and the associated saddle point equations analogous to Eqs.\eqref{seff} and \eqref{aa3}. In the limit $q\gg 1$, the saddle point equations can be consistently expanded in terms of 
\begin{align}
G(\tau) \underset{q\gg 1}{=} \frac{1}{2}\sgn(\tau) \left[1+q^{-1}g(\tau)+O(q^{-2})\right]
\end{align}
\noindent yielding a non-linear boundary value problem for $g$,
\begin{align}\label{eq:little_g}
\partial_{\theta}^{2}g = 2 (\beta  J)^2e^{g(\theta)} +(\beta \kappa)^2,
\end{align}
\noindent with $\theta = \tau/\beta \in (0, 1)$. The retarded kernel Eq.\eqref{eq:largeq:retarded_kernel} is thus obtained by analytical continuation of the resulting $G(\tau)$ and, as mentioned before, for $q\gg1$ it is given by a simpler expression,
\begin{align}
K_R(t_1,t_2,t_3,t_4)=\theta(t_{13})\theta(t_{24})\left(2 J^2 e^{g(\tau=it_{34}+\beta/2)}+q^{-1}\kappa^2 \right)\,,
\end{align}
where $g(\tau)$ is the solution of Eq.\eqref{eq:little_g} and is given explicitly as a power series in $\kappa/J\ll1$ in the Supplemental Material, Sec. C.
We again use the ansatz ${\cal F}(t_1,t_2)=e^{\lambda_L(t_1+t_2)/2}f(t_{12})$ in order to rewrite the eigenvalue problem in Eq.\eqref{eq:otos} as a Schr\"{o}dinger equation for $f(t_{12})$. The eigenstates of the resulting equation are found perturbatively in $\kappa/J\ll1$, giving a correction ${O}(\kappa^2)$ to the Lyapunov exponent. This correction is given in terms of an integral that, for low temperature $\beta{J}\gg1$, is approximated by 
\begin{align}\label{eq:Lyapunov_kappa}
\frac{\beta\lambda_L}{2\pi}\bigg\rvert_{\substack{q\gg1\\ \kappa\ll J} }&{=}1{-}\frac{(\beta \kappa )^2}{\pi^2}\left[{1\over 72}+\frac{19-18\log\pi}{36\beta J}+O\left({1\over(\beta J)^{2}}\right)\right].
\end{align}
The transition occurs when $\lambda_L = 0$, which leads to a $\beta$-dependent critical $\kappa = \kappa_c$. For instance, $\kappa_c(\beta{=}133)\sim 0.2J$ and $\kappa_c(\beta{=}53)\sim 0.5J$, which is in good agreement (see Fig. \ref{fig:Lyapunov}) with numerical results for $q=4$.
 We refer to Sec. C of the Supplemental Material for additional details.

Finally, we note these types of transitions are generic \cite{stasio1995}, so it would be interesting to identify their gravity dual. We speculate with the possibility that the gravity dual of the transition studied in this Letter is a Hawking-Page transition where the black hole and thermal gas phases correspond to the chaotic and integrable phase, respectively. 
In conclusion, we have found that the SYK model perturbed by a random one-body term is still chaotic in the limit of  sufficiently high temperature or weak perturbation. However, for a given strength of the perturbation, the system undergoes a chaotic-to-integrable transition for sufficiently low temperatures which may have a gravity dual interpretation.  

 {\it Note added}: Close to completion of this work, we became
 	aware of three papers \cite{song2017,chen2017,eberlein2017} that study a somehow similar generalized SYK model though the focus of these papers is rather different. 
 	Ref. \cite{eberlein2017} studies quantum quenches while the other two investigate a two fermion species generalization of the model in which a transition occurs by tuning the number of fermions.  
\acknowledgments
We thank D. Anninos, S. Banerjee, and P. Sabella-Garnier for illuminating discussions.
Part of the computation in this Letter has been done using the facilities of the Supercomputer Center, the Institute for Solid State Physics, the University of Tokyo.
A. M. G. acknowledges partial financial support from a QuantEmX
grant from ICAM and the Gordon and Betty Moore
Foundation through Grant GBMF5305.
The work of M. T. was partially supported 
by Grants-in-Aid No. JP26870284 and No. JP17K17822 from JSPS of Japan. A. R. B. is funded through 
a research programme of the Foundation for Fundamental Research on Matter (FOM),
which is part of the Netherlands Organisation for Scientific Research (NWO). B. L. is supported by a CAPES/COT grant No. 11469/13-17.

\appendix\label{apptp}
\section{Appendix A: Numerical Thermodynamic properties}
\begin{figure}
	\hspace{-5mm}
	\resizebox{0.5\textwidth}{!}{\includegraphics{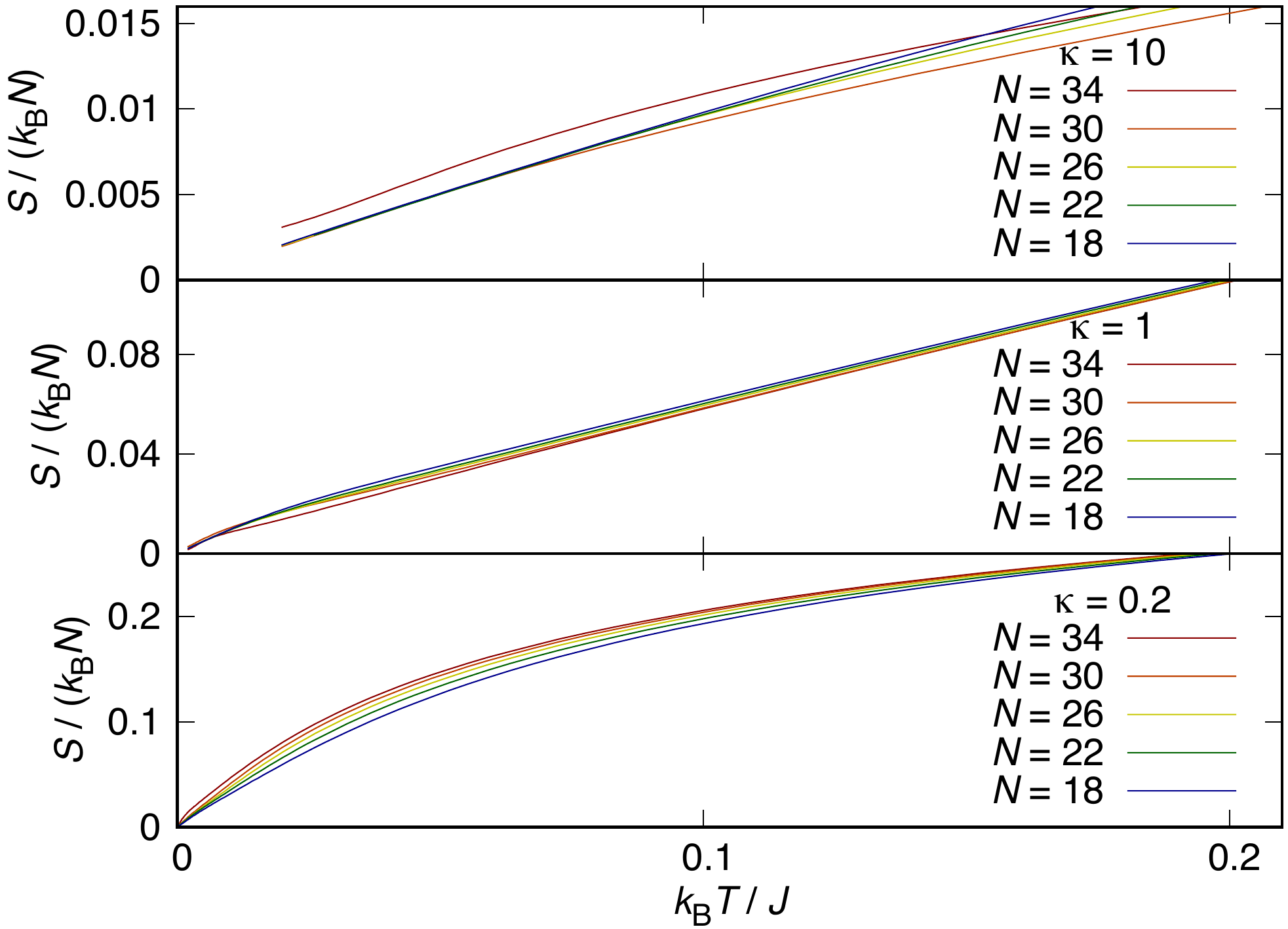}}
	\caption{Entropy $S$ as a function of temperature $T$ from the exact diagonalisation of Eq.(\ref{hami}) for different $N$'s and $\kappa$ where $k_B$ stands for the Boltzmann constant. The number of eigenvalues employed is mentioned in the main text. The weak dependence on $N$ is consistent with a vanishing zero temperature entropy. }
	\label{fig4}%
\end{figure}

In this appendix we present explicit results for the low temperature limit of the entropy and the specific heat that, for space limitations, could not be included in the main text.
The zero temperature limit of the entropy, $s_0$, was obtained by exact diagonalisation of the Hamiltonian Eq.(\ref{hami}) for different $N$ and $\kappa$.
We note that for any finite $N$ the entropy will vanish in the $T \to 0$ limit \footnote{The reason is simply that for finite $N$ there is a gap of the order $2^{-N}$ so that degeneracy is not exact.}.
Therefore, to justify a zero entropy in this limit it is necessary to extrapolate the finite $N$ numerical results to the $N \to \infty$ limit. However, this does not seem strictly necessary as the $N$ dependence, depicted in Fig.~\ref{fig4}, is very weak, especially for $\kappa \geq 1$. A simple extrapolation of the curves for larger $N$ to the $T=0$ limit leads to a zero-temperature entropy which is smaller than $10^{-3}$ for all $\kappa$'s. These results are fully consistent with the $N\to\infty$ results, which have been obtained by fitting ${\log Z\over N}=-E_0\beta +s_0+{c\over 2\beta}+{c_1\over \beta^2}+{c_2\over \beta^3}$ for $J = K = 1$. In order to obtain ${\log Z\over N}$ we solve Eq.~(\ref{aa3}) as described previously in Refs. \cite{maldacena2016,davison2017}. We use a Fast Fourier transform to switch between frequency and time domains and solve iteratively until convergence. For $\omega_n=2\pi T(n+1/2)$ we take $- N_\omega/2 < n <N_\omega/2-1$, with $N_\omega=2^{24}$ and $N_t=4N_\omega$ points in the frequency and time domains.  

We now move to the study of the low temperature limit of the specific heat $C(T)$ per Majorana obtained by exact diagonalisation. For that purpose we employ the following thermodynamic expression:
\be
C(T) = \left \langle \frac {1}{NZ} \sum_{k} \frac{(E_{k} - \bar E )^2}{T^2}e^{-\beta E_{k}} \right \rangle\,,
\label{c2}
\ee
where $Z$ is the partition function, $\langle \ldots \rangle$ stands for ensemble average and $k$ labels the eigenvalues for a given disorder realisation with average $\bar E$.

\begin{figure}[H]
	\centering
	\resizebox{0.46\textwidth}{!}{\includegraphics{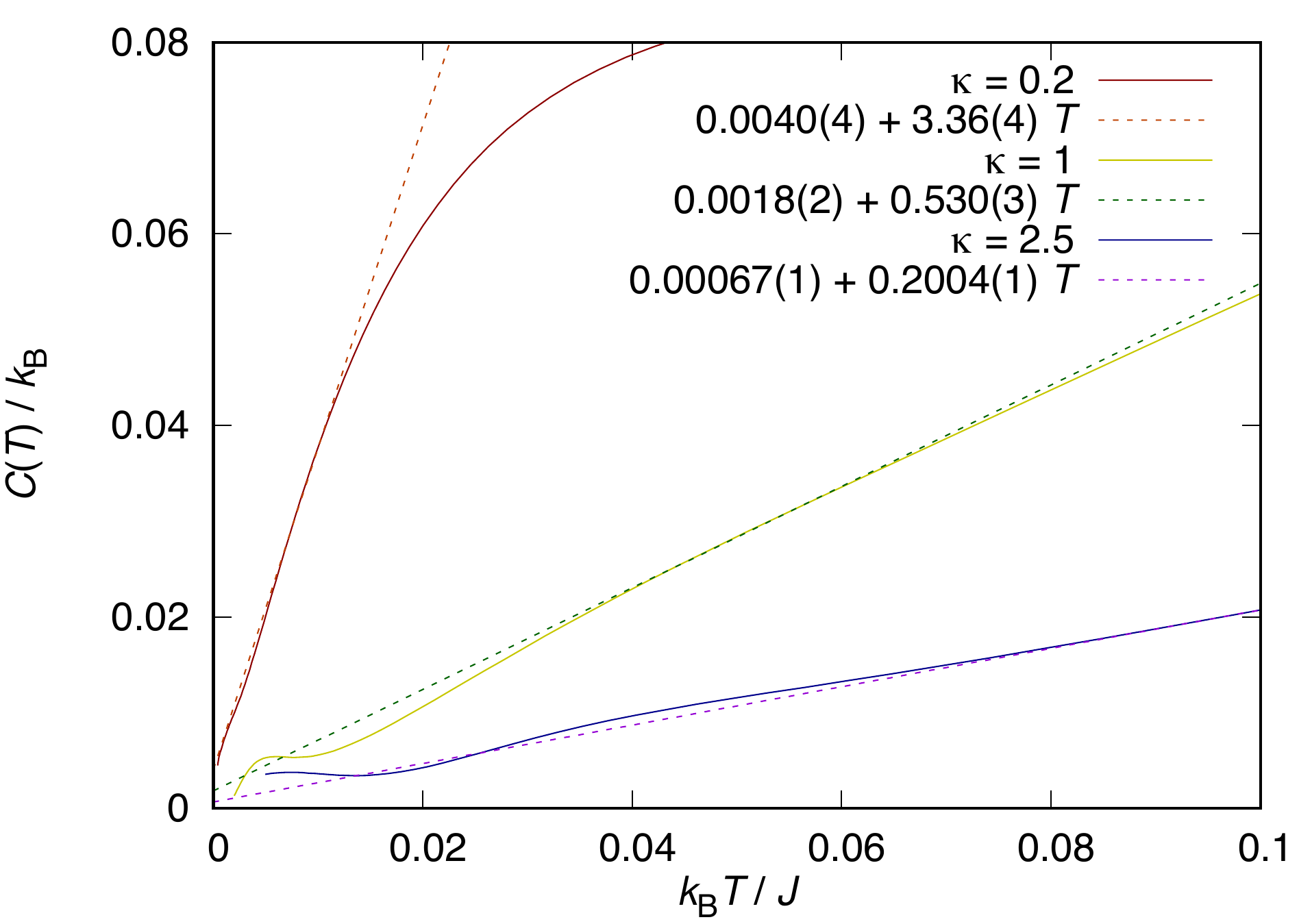}}
	\vspace{-5mm}
	\caption{Specific heat $C(T)$ as a function of temperature, in units of $J/\kappa_B$, obtained from Eq.(\ref{c2}) and the exact diagonalisation of Eq.(\ref{hami}) for $N = 34$ and different $\kappa$'s. The specific heat is clearly linear in the low temperature limit with a slope that it is close to the $\kappa = 0$ prediction $c =\pi/6$ for any $\kappa \leq 1$. }
	\label{fig5} 
\end{figure}

\begin{figure}[H]
	\centering
	\resizebox{0.46\textwidth}{!}{\hspace{-2mm}\includegraphics{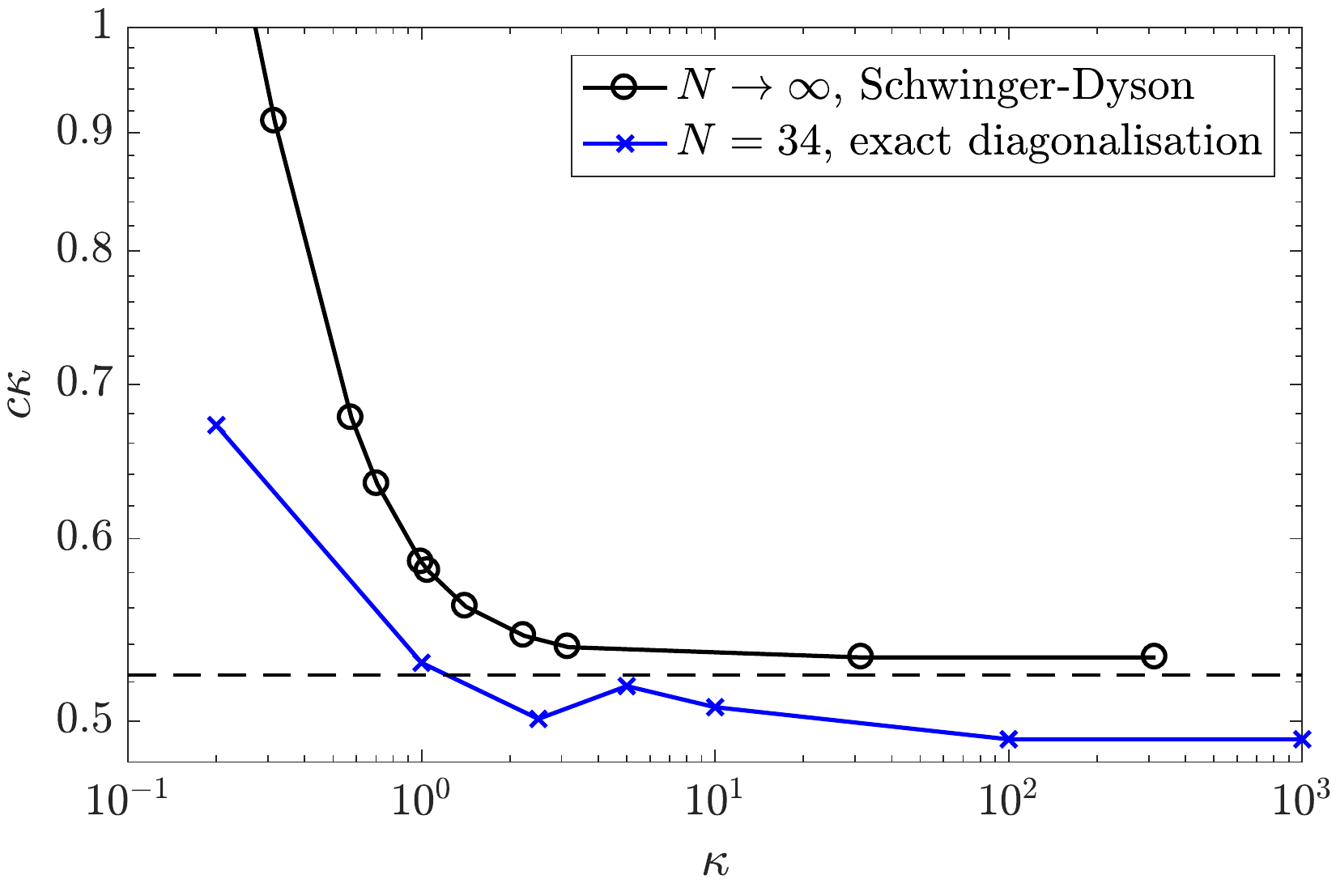}}
	\vspace{-3mm}
	\caption{Specific heat coefficient $c$ as a function of $\kappa$. The exact diagonalisation result, shown in blue crosses, is the slope of the fitting curve in Fig.~\ref{fig5}. The black dots are obtained from fitting $-\beta {F\over N}={\log Z\over N}$ obtained from the numerical solution of the large-$N$ saddle point equations Eq.(\ref{aa3}). The dashed line corresponds to $c=\pi/(6\kappa)$, the analytical value when the Hamiltonian only contains the random-mass term  \cite{maldacena2016}. Differences between the two results are consistent with $1/N \sim 0.03$ corrections only retained in the exact diagonalisation result.}
	\label{fig:c_compare} 
\end{figure}

We carry out quenched averages, namely, the specific heat is computed separately for each disorder realisation. The final specific heat is the arithmetic average over all disorder realisations. We then fit the low temperature limit by a low order polynomial in temperature. The coefficient of the linear term is the specific heat coefficient, $c$, which in units of $N$ and with the coupling constant set to one, is $\pi/(6\kappa)$ for $\kappa \to \infty$ and $\approx 0.4$ for the unperturbed two-body SYK model ($\kappa=0$) \cite{maldacena2016,garcia2016}. As shown in Fig.~\ref{fig5}, for large $\kappa$, $c$ is indeed very close to $\pi/(6\kappa)$. Only for $\kappa \ll 1$ we observe a moderate increase of $c$. 
This is a further confirmation that the ground state and lowest energy excitations of the Hamiltonian Eq.(\ref{hami}) are mostly controlled by the random-mass term.
We cannot study arbitrarily small $\kappa$ because it would require to reach very low temperatures that compromise the numerical accuracy of the results. We have checked that the obtained specific heat coefficient $c$ is robust to 
changes in the fitting interval. We have observed that for $N \leq 30$ the value of $c$ does not have a monotonic dependence on $N$ which makes difficult to carry out a finite-size scaling analysis. For that reason, unless otherwise stated, the fitting is restricted to the largest $N = 34$ that can be reached numerically. Finally, in Fig.~\ref{fig:c_compare}, we compare the specific heat coefficient $c$ obtained from the large-$N$ fitting of $\log Z\over N$, as explained above with the exact diagonalisation result given in Fig.~\ref{fig5}. Deviations are consistent with $1/N$ corrections only taken into account in the latter.

\section{Appendix B: Analytical Low Temperature Thermodynamic properties}
\label{appendix:B}
We carry out an analytical calculation of the low temperature thermodynamic properties of the generalised one+two-body SYK model in the limit of large $N$. Since the exact solution for the one-body Hamiltonian is known analytically, it will be convenient to develop the perturbative approach around this exact ground state.
For this purpose, it will be convenient to consider a rescaled model given by 
\begin{align}\label{eq:SYK4_hami}
\tilde{H} = \frac{i}{2!} \sum\limits_{i, j = 1}^{N}	\kappa_{ij}\chi_i\chi_j + \frac{\tilde{\kappa}}{4!}\sum\limits_{i,j,k,l=1}^{N}J_{ijkl}\ \chi_i \chi_j \chi_k \chi_l\,,
\end{align}
where the standard deviation of the random couplings $J_{ijkl}$ and $\kappa_{ij}$ are given by $\frac{\sqrt{6}\tilde{\kappa}}{N^{3/2}}$ and $\frac{1}{\sqrt{N}}$ respectively \footnote{Denoting the Hamiltonian in Eq.\eqref{hami} by $H(\kappa)$, we have that $\tilde{H}(\tilde{\kappa}) = \tilde{\kappa}^{-1}H(\tilde{\kappa}^{-1})$ with $J=1$.}. The disorder averaged effective action associated with the rescaled Hamiltonian is given by
\begin{align}
\label{appB:eq:effective}
\tilde{S}_\mathrm{eff}&=-\frac{1}{2}\Tr \log( \partial_\tau-\Sigma)+{1\over2}\int \dd \tau\dd\tau'\left[G(\tau,\tau')\Sigma(\tau,\tau')\right. \nonumber\\
&\left.-{\tilde{\kappa}^2\over 4!}G(\tau,\tau')^4-{1\over 2}G(\tau,\tau')^2\right]\,,
\end{align}
where $G(\tau,\tau') \equiv -{1\over N}\sum_{i=1}^N\langle T \chi_i(\tau)\chi_i(\tau')\rangle$ is the Euclidean many-body correlator.

 We will proceed in two steps, first we find a perturbative solution of the rescaled SD equations and then we compute the low temperature limit of the free energy by substitution of the obtained Green's function in the action.

{\it Perturbative solution of the SD equation.-} The SD equations derived from Eq.\eqref{appB:eq:effective} can be written in Fourier space as
\begin{align}
\label{eq:SP_fourier}
G(\omega_n)^2+\left[i\omega_n+{\tilde{\kappa}}^2~\sigma(\omega_n)\right]G(\omega_n)+1=0,
\end{align}
\noindent where we have defined $\sigma(\omega) = \int\dd\tau ~ e^{i\omega_n\tau}G^{\beta}(\tau)^3$. For ${\tilde{\kappa}}=0$, the system is noninteracting and Eq.\eqref{eq:SP_fourier} simplifies to an algebraic quadratic equation in $G$, which is solved exactly by
\begin{align}
\label{eq:zero_order}
G_{0}(\omega_n) = \frac{-i\omega_n+i\sgn(\omega_n)\sqrt{4+\omega_n^2}}{2}.
\end{align}
The subindex `$0$' in $G_{0}(\omega_n) $ indicates it is the solution to the $\tilde \kappa=0$ noninteracting limit.
For $\tilde\kappa>0$, Eq.\eqref{eq:SP_fourier} is an integral equation, and cannot be solved exactly. We proceed with a perturbative solution in the limit $\tilde\kappa \ll 1$. Let
\begin{align}
\label{eq:perturbative}
G(\omega_n) \underset{{\tilde{\kappa}}\ll 1}{=} G_{0}(\omega_n) + {\tilde{\kappa}}^2~ g(\omega_n) + O({\tilde{\kappa}}^4).
\end{align}
Inserting in Eq.\eqref{eq:SP_fourier} and expanding in ${\tilde{\kappa}}$, we find
\begin{align}
\label{eq:correction}
g(\omega_n) = -\frac{G_0(\omega_n)}{i\omega_n+2G_{0}(\omega_n)}	\sigma_0(\omega_n)
\end{align}
\noindent where $\sigma_{0}(\omega) = \int\dd\tau ~ e^{i\omega\tau}G_0(\tau)^3$ acts as a source for the first order correction. The prefactor can be written exactly,
\begin{align}
\label{eq:prefactor}
\frac{G_0(\omega_n)}{i\omega_n+2G_{0}(\omega_n)}=\frac{1}{2}\left[1-\frac{\omega_n~\sgn(\omega_n)}{\sqrt{4+\omega_n^2}}\right].
\end{align}
However $\sigma_0(\omega_n)$ requires more effort since we first need $G_0^{\beta}(\tau)$. Note that the only temperature dependence of $G_0(\omega_n)$ is through the Matsubara frequencies, and thus disappear upon analytic continuation $i\omega_n\to \epsilon+i0^+$. This is expected since ${\tilde{\kappa}}=0$ is essentially a noninteracting system, and the spectral density should not be temperature dependent. However, for any ${\tilde{\kappa}}>0$ the system is interacting, and we do expect non-trivial temperature dependence in $G(\omega_n)$. As we will see below, this comes exactly from the non-linearity induced by the $\sigma_0$ contribution.

{\it Finite temperature.-} We analytically continue $i\omega_n \to \epsilon+i0^+$ Eq.\eqref{eq:zero_order} to get the zeroth order (in $\tilde \kappa$) retarded Green's function which corresponds to the noninteracting limit
\begin{align}
\label{eq:retardedsyk2}
G^{R}_0(\epsilon)=\frac{-\epsilon+i\sqrt{4-\epsilon^2}}{2}\,,
\end{align}
Note that, as discussed above, temperature dependence has disappeared. The zeroth order spectral function is 
\begin{align}
\label{eq:zero_specdensity}
	\rho_0(\epsilon) &= 2\text{Im }G^R_0(\epsilon) =\sqrt{4-\epsilon^2} \mathbb{I}_{[-2,2]}(\epsilon),
\end{align}
\noindent where $\mathbb{I}_A(x)$ is the characteristic function of the set $A$. Note this is the well-known Wigner semicircle distribution for the spectral density of a random matrix. It should come at no surprise since for ${\tilde{\kappa}}=0$ the Hamiltonian of the system is a sparse random matrix. The spectral function allow us to analytically continue $G_0(\omega_n)$ to the whole complex plane as
\begin{align}
G_0(z) = \int_{\mathbb{R}}\frac{\dd \lambda}{2\pi} \frac{\rho(\lambda)}{\lambda-z},	
\end{align}
\noindent and provide a useful way of getting the different Green's functions. To get the finite temperature Matsubara Green's function, we set $z=i\omega_n$ and sum over the  frequencies $G^{\beta}(\tau) = \sum\limits_{n\in\mathbb{Z}}e^{i\omega_n \tau}G(i\omega_n)$, as usual in the Matsubara formalism: 
\begin{align}
\label{eq:matsubara}
G_0^{\beta}(\tau) = \int_{\mathbb{R}}\frac{\dd\lambda}{2\pi} \frac{\rho(\lambda)e^{-\lambda \tau}}{1+e^{-\beta \lambda}}\,.
\end{align}
In principle we have everything we need to compute $\sigma_0(\omega_n)$ and thus the first order correction $g(\omega_n)$. However, in the lack of a closed expression, we proceed with a low temperature expansion of Eq.\eqref{eq:matsubara}. More, precisely we expand the integrand of Eq.\eqref{eq:matsubara}, where  $\rho(\lambda)  = \rho_0(\lambda)$ is given in Eq.\eqref{eq:zero_specdensity}, for large $\beta$. We then perform the integration order by order giving
\begin{widetext}
\begin{align}
G_{0}^{\beta}(\tau) \underset{\beta \gg 1}=&\left(-\frac{5 \pi ^8}{32768 \beta ^9}+\frac{\pi ^6}{1024 \beta ^7}-\frac{\pi ^4}{128 \beta ^5}+\frac{\pi ^2}{8 \beta ^3}+\frac{1}{\beta }\right) \csc{\pi\theta}	+\left(\frac{1025 \pi ^8}{4096 \beta ^9}-\frac{91 \pi ^6}{512 \beta ^7}+\frac{5 \pi ^4}{32 \beta ^5}-\frac{\pi ^2}{4 \beta ^3}\right)\csc^3{\pi\theta}\notag\\
&+\left(-\frac{7245 \pi ^8}{2048 \beta ^9}+\frac{105 \pi ^6}{128 \beta ^7}-\frac{3 \pi ^4}{16 \beta ^5}\right)\csc^5{\pi\theta}+\left(\frac{4725 \pi ^8}{512 \beta ^9}-\frac{45 \pi ^6}{64 \beta ^7}\right)\csc^7{\pi\theta}-\frac{1575 \pi ^8}{256 \beta ^9}\csc^9{\pi\theta}+O(\beta^{-11}),
\end{align}
\end{widetext}
\noindent where we define $\theta = \tau/\beta \in (0,1)$. We now expand $G_0^{\beta}(\tau)^3$ in $1/\beta$ and perform the Fourier transform to get $\sigma_0(\omega_n) = \int_{0}^{\beta}\dd\tau ~ e^{i\omega_n \tau}G^{\beta}_0(\tau)^3$. Inserting this into Eq.\eqref{eq:correction}, together with Eq.\eqref{eq:prefactor}, we get an expression for $g(\omega_n)$ that should be added to the zeroth order solution Eq.\eqref{eq:zero_order} (see Eq.\eqref{eq:perturbative}). Analytically continuing this expression $i\omega_n \to \epsilon+i0^+$ and expanding in low frequencies gives an approximation for the analytic continuation of Eq.\eqref{eq:perturbative} $G^R(\epsilon)=G(i\omega_n\to \epsilon)$:
\begin{widetext}
\begin{align}
\label{eq:retarded_first}
G^{R}(\epsilon) \underset{\epsilon\ll 1}{=}	&\epsilon\left[\left(-\frac{3 \pi ^7}{8192 \beta ^8}-\frac{3 \pi ^3}{128 \beta ^4}+\frac{\pi }{8 \beta ^2}\right) {\tilde{\kappa}}^2 -\frac{1}{2}\right]+\tilde{\kappa}^2\epsilon^3\left(-\frac{3 \pi ^7}{65536 \beta ^8}-\frac{913 \pi ^5}{645120 \beta ^6}-\frac{17 \pi ^3}{3840 \beta ^4}-\frac{\pi }{64 \beta ^2}+\frac{1}{8 \pi }\right)\notag\\
&+\tilde{\kappa}^2\epsilon^7\left(-\frac{9 \pi ^7}{1048576 \beta ^8}-\frac{913 \pi ^5}{5160960 \beta ^6}-\frac{41 \pi ^3}{20480 \beta ^4}-\frac{\pi }{384 \beta ^2}+\frac{1}{128 \pi
   }\right)\notag\\
   & +\tilde{\kappa}^2\epsilon^7\left(-\frac{15 \pi ^7}{8388608 \beta ^8}-\frac{913 \pi ^5}{27525120 \beta ^6}-\frac{121 \pi ^3}{393216 \beta ^4}-\frac{17 \pi }{30720 \beta ^2}+\frac{7}{3840
   \pi }\right)\notag\\
   &+i\epsilon^2 \left[\left(\frac{913 \pi ^5}{322560 \beta ^6}+\frac{23 \pi ^3}{7680 \beta ^4}+\frac{ \pi }{16 \beta ^2}-\frac{1}{4 \pi }\right) \tilde{\kappa}^2 -\frac{1}{8}\right]+i\epsilon^4\left[\left(\frac{311 \pi ^3}{122880 \beta ^4}+\frac{5 \pi }{1536 \beta ^2}+\frac{1}{64 \pi }\right) \tilde{\kappa}^2 -\frac{1}{128}\right]\notag\\
   &+i\epsilon^6\left[\left(\frac{7 \pi }{15360 \beta ^2}+\frac{1}{3840 \pi }\right) \tilde{\kappa}^2 -\frac{1}{1024}\right]+i\epsilon^8\left[\frac{59 \tilde{\kappa}^2 }{2580480 \pi }-\frac{5}{32768}\right]+O(\epsilon^9,\tilde{\kappa}^4).
\end{align}
\end{widetext}
\noindent where we have conveniently separated the real and imaginary parts. The real part of the retarded Green's function is odd, while the imaginary part which gives the spectral density is even. Note that the effect of interactions is mainly to renormalise the coefficients of the free $\tilde{\kappa}=0$ model, with both zero and finite temperature contributions. This reflects the fact that the interactions are irrelevant compared to the hopping term. At zero temperature $\beta=\infty$, corrections only appear at order $\epsilon^2$. This means that the low frequency behaviour of the ground state of the system is unchanged.

{\it Low temperature expansion of the free energy.-} To compute the free energy density of the model, we have to insert the finite temperature saddle point solution we obtained in the sections above into the effective action Eq.\eqref{appB:eq:effective}. As mentioned earlier, we expect the low temperature expansion to be given by
\begin{align}
\label{eq:free_en_expansion}
-\frac{\beta F}{N} = -\beta E_0 + s_0 + \frac{c}{2\beta} + O(\beta^{-2}).	
\end{align}
\noindent where $E_0$ is the ground state energy density, $s_0$ the zero temperature entropy density and $c$ the specific heat coefficient. Our aim in this section is to compute these coefficients for small $\tilde{\kappa}$. One can check that the integral terms in Eq.\eqref{seff} do not contribute to these lowest order coefficients, and the leading contributions come from the $\text{Tr}\log$ term. By definition, it is given by
\begin{align}
\text{Tr}\log G = \beta\sum\limits_{n\in\mathbb{Z}} \log{G(i\omega_n)} e^{i\omega_n 0^+}.
\end{align}
We can write this sum over Matsubara frequencies as the residue of the complex integral $\int_\gamma\frac{\dd z}{2\pi i}\log G(z)n(z) e^{z0^+}$ where $n(z) = (1+e^{\beta z})^{-1}$ and the contour $\gamma$ englobes the poles at $z=i\omega_n$. The integrand also has a branch cut along the real axis. Since the integral decay at infinity, we can deform $\gamma$ to wrap around the branch cut in the real axis. This leads to
\begin{align}
\label{eq:trlog}
\frac{1}{2}\text{Tr}\log G &= \beta\int_{\gamma} \frac{\dd z}{2\pi i} \log{G(z)}n(z) e^{z 0^{+}}\notag\\
&= \beta\int_{\mathbb{R}}\frac{\dd \epsilon}{2\pi i} n(\epsilon)\left[\log{G(\epsilon+i0^+) - \log{G(\epsilon-i0^+)}} \right]\notag\\
&=\beta\int_{\mathbb{R}}\frac{\dd \epsilon}{\pi} n(\epsilon)\text{Im}\log{G^R(\epsilon)} = \beta\int_{\mathbb{R}}\frac{\dd \epsilon}{\pi} \frac{\text{Arg }{G^R(\epsilon)}}{1+e^{\beta \epsilon}}\,,
\end{align}
where $G^R(\epsilon) = G(i\omega_n\to \epsilon+i0^+)$.
For $\tilde{\kappa} = 0$, $G^R(\epsilon)$ is simply given by Eq.\eqref{eq:retardedsyk2}. Thus,
\begin{align}
\text{Arg }{G_{0}^{R}} 	=\begin{cases}
\frac{\pi}{2}-\tan^{-1}\left(\frac{-\epsilon}{\sqrt{4-\epsilon^2}}\right) &\text{ for } |\epsilon|<2 ,\\ 
\pi & \text{ if } \epsilon\geq 2,\\
0 & \text{ if } \epsilon\leq -2 ,
\end{cases}
\end{align}
\noindent giving
\begin{align}
\label{eq:trlogsyk2}
\frac{1}{2}\text{Tr}\log{G_0} &= \beta\int_{-2}^{2}\frac{\dd \epsilon}{\pi} \frac{\frac{\pi}{2}-\tan^{-1}\left(\frac{-\epsilon}{\sqrt{4-\epsilon^2}}\right)}{1+e^{\beta \epsilon}}+\beta \int_{2}^{\infty}\frac{\dd\epsilon}{1+e^{\beta\epsilon}}\notag\\
&\underset{\beta\gg 1}{=}\beta(\pi-1)+\log\left(1+e^{-2\beta}\right)+\frac{\pi}{12\beta}+O(\beta^{-3}),
\end{align}
\noindent where in the last equality we used the Sommerfeld expansion for the Fermi-Dirac distribution $n(\epsilon)\underset{\beta\gg 1}=\theta(-\epsilon)-\frac{\pi^2}{6\beta^2}\delta'(\epsilon)+O(\beta^{-4})$. The $\log$ term is exponentially decaying, and does not contribute in Eq.\eqref{eq:free_en_expansion}. For $\tilde{\kappa} = 0$ we restore the units by taking the coupling $\kappa_{ij}$ to have a standard deviation parametrized by a scale $K$: $K/\sqrt{N}$. Therefore, we conclude that for $\kappa=0$ we have $E_0 = (1-\pi)~K N$, $s_0=0$ and $c=\frac{\pi}{6} \frac{N}{K}$.
These values are consistent with the previous results in the literature for the SYK model at $q=2$ \cite{maldacena2016}. Note that the term proportional to $\beta^{-1}$ in the low temperature expansion only depends on derivatives of the argument evaluated at $\epsilon =0$. Thus for $\tilde{\kappa}=0$ the specific heat coefficient only depends on the low frequency properties of the retarded Green's function. 

\begin{figure}[h!]
\centering
	\hspace{-5mm}
		\resizebox{0.49\textwidth}{!}{\includegraphics{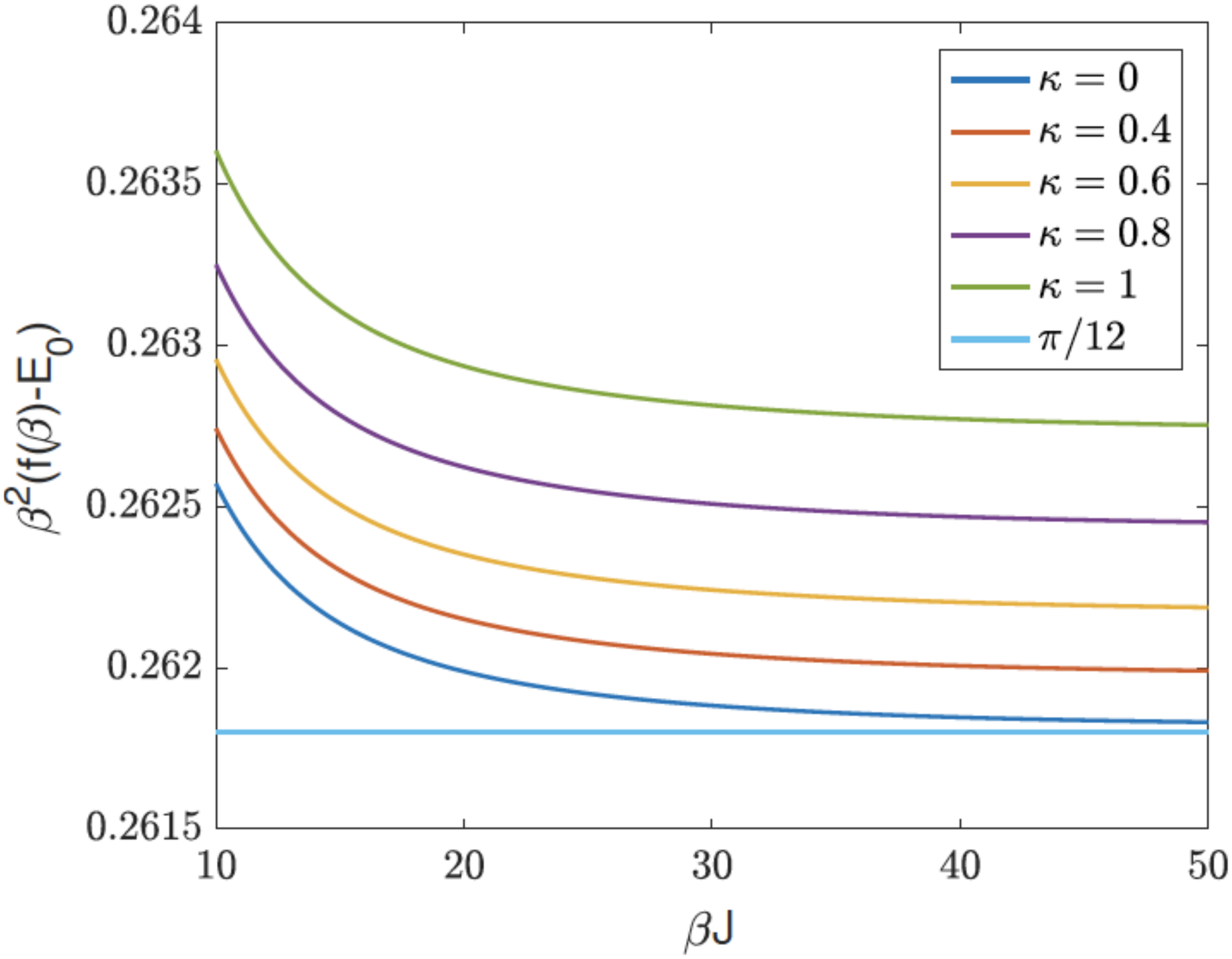}}
	\vspace{-4mm}
\caption{Numerical integration of $\beta^2(F/N-E_0)$, which asymptotes to $c/2$ at low temperatures.}
\label{fig:1}
\end{figure}

\begin{figure}[h!]
\centering
	\hspace{-5mm}
		\resizebox{0.49\textwidth}{!}{\includegraphics{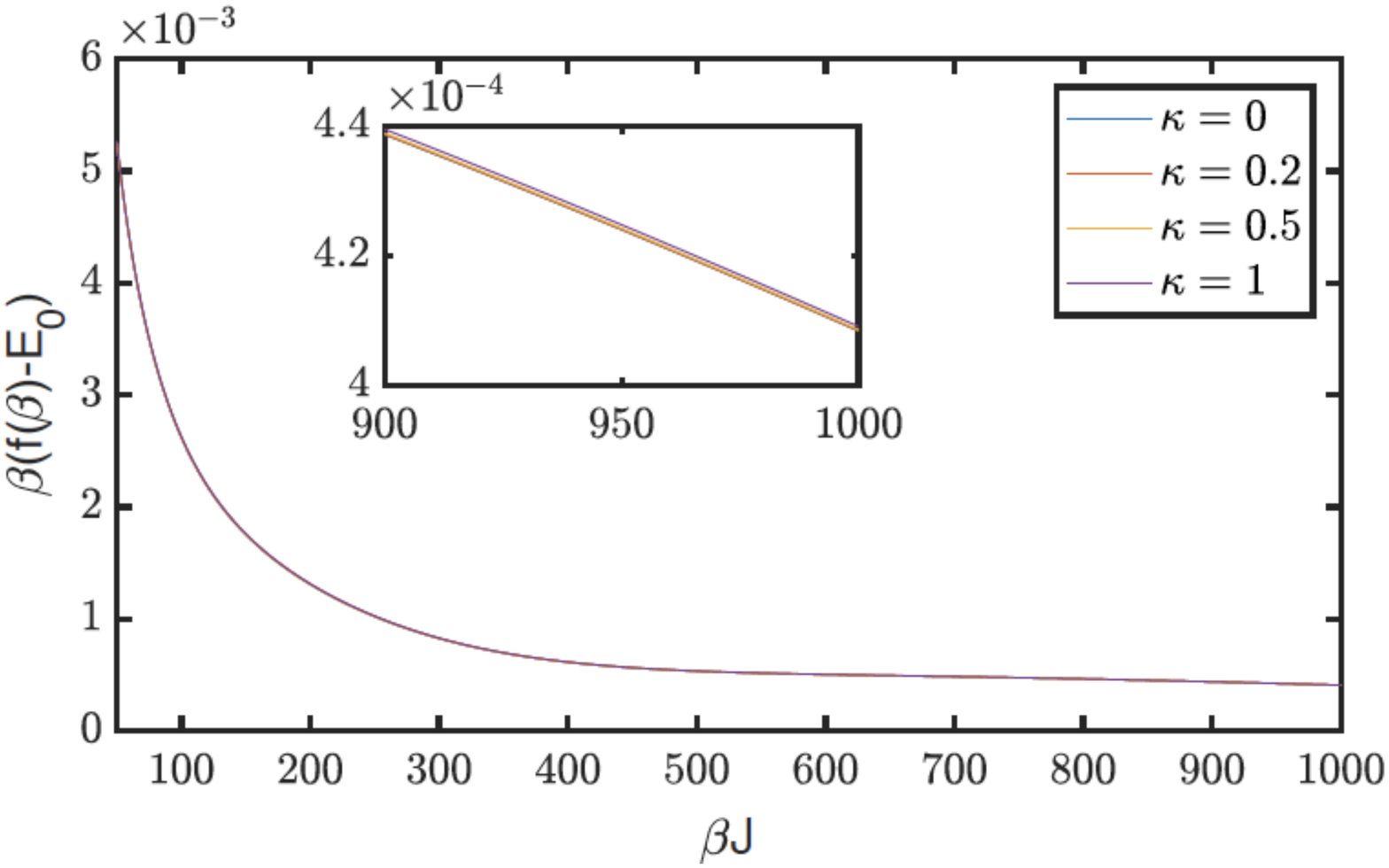}}
	\vspace{-10mm}
\caption{Numerical integration of $\beta(F/N-E_0)$, which asymptotes to $s_0$ at low temperatures.}
\label{fig:3}
\end{figure}

For $\tilde{\kappa}>0$, we only have access to the IR low frequency result in Eq.\eqref{eq:retarded_first}. Naively inserting this in Eq.\eqref{eq:trlog}  leads to UV divergences due to unboundedness of the integrals. Note this was not needed for $\tilde{\kappa}=0$ since the compact support of the spectral function comes naturally from the exact solution. But UV divergences are expected since the perturbative IR solution does not hold for large frequencies. We therefore introduce an UV cutoff $\lambda$ to regularise the integrals. We will show that this leads to cutoff dependent ground state energy, but cutoff independent entropy and specific heat. This is expected given that, as discussed above, the specific heat is determined by the IR behaviour of $G^R$. Although there is some freedom in the choice of $\lambda$, it is largely constrained by non-perturbative properties of the spectral function $\rho(\epsilon) = 2\text{Im }G_R$, which must satisfy the sum rule $\int \frac{\dd\epsilon}{2\pi}\rho(\epsilon) = 1$ and positivity $\rho(\epsilon)\geq 0$. These constraints give us a consistent way to fix the cutoff: we choose $\lambda$ such that $\int_{-\lambda}^{\lambda} \frac{\dd\epsilon}{2\pi}\rho(\epsilon) = 1$ \footnote{Note that since $\rho$ is even, without loss of generality we can take a symmetric interval $(-\lambda,\lambda)$.}. This ensures that $|\epsilon|< \lambda$ saturates the spectral weight, and therefore for $|\epsilon|\geq \lambda$ the perturbative solution breaks down. It is easy to check that this choice automatically satisfies positivity of $\rho$. With this discussion in mind, we take
\begin{align}
\label{eq:phaseshift}
\text{Arg }{G^{R}} =\begin{cases}
\frac{\pi}{2}-a(\epsilon,\tilde{\kappa},\beta) &\text{ for } |\epsilon|<\lambda ,\\ 
\pi & \text{ if } \epsilon\geq \lambda,\\
0 & \text{ if } \epsilon\leq -\lambda ,
\end{cases}
\end{align}
\noindent where we define $a(\epsilon,\tilde{\kappa},\beta) = \tan^{-1}\left(\frac{\text{Re }{G^R}}{\text{Im }{G^R}}\right)$ with $G^{R}$ given in Eq.\eqref{eq:retarded_first}. As in Eq.\eqref{eq:trlogsyk2}, the integral over $(\lambda,\infty)$ gives an exponentially decaying term that does not contribute to the thermodynamic coefficients. The remaining integral over $(-
\lambda,\lambda)$ can be integrated numerically using the cutoff estimated from the sum-rule for $\rho$.\footnote{We have also  checked numerically that, as long and positivity of $\rho$ is respected, changes in the cutoff do not affect the results for the thermodynamic coefficients $c$ and $s_0$.}. To extract the specific heat coefficient, we subtract the zero temperature result and multiply by a factor $\beta^2$. According to Eq.\eqref{eq:free_en_expansion}, the result should asymptote to $c/2$ as $\beta\to\infty$. These are shown in Fig.~\ref{fig:1} and Fig.~\ref{fig:3} respectively.

\begin{figure}[h!]
\centering
	\hspace{-5mm}
	\resizebox{0.49\textwidth}{!}{\includegraphics{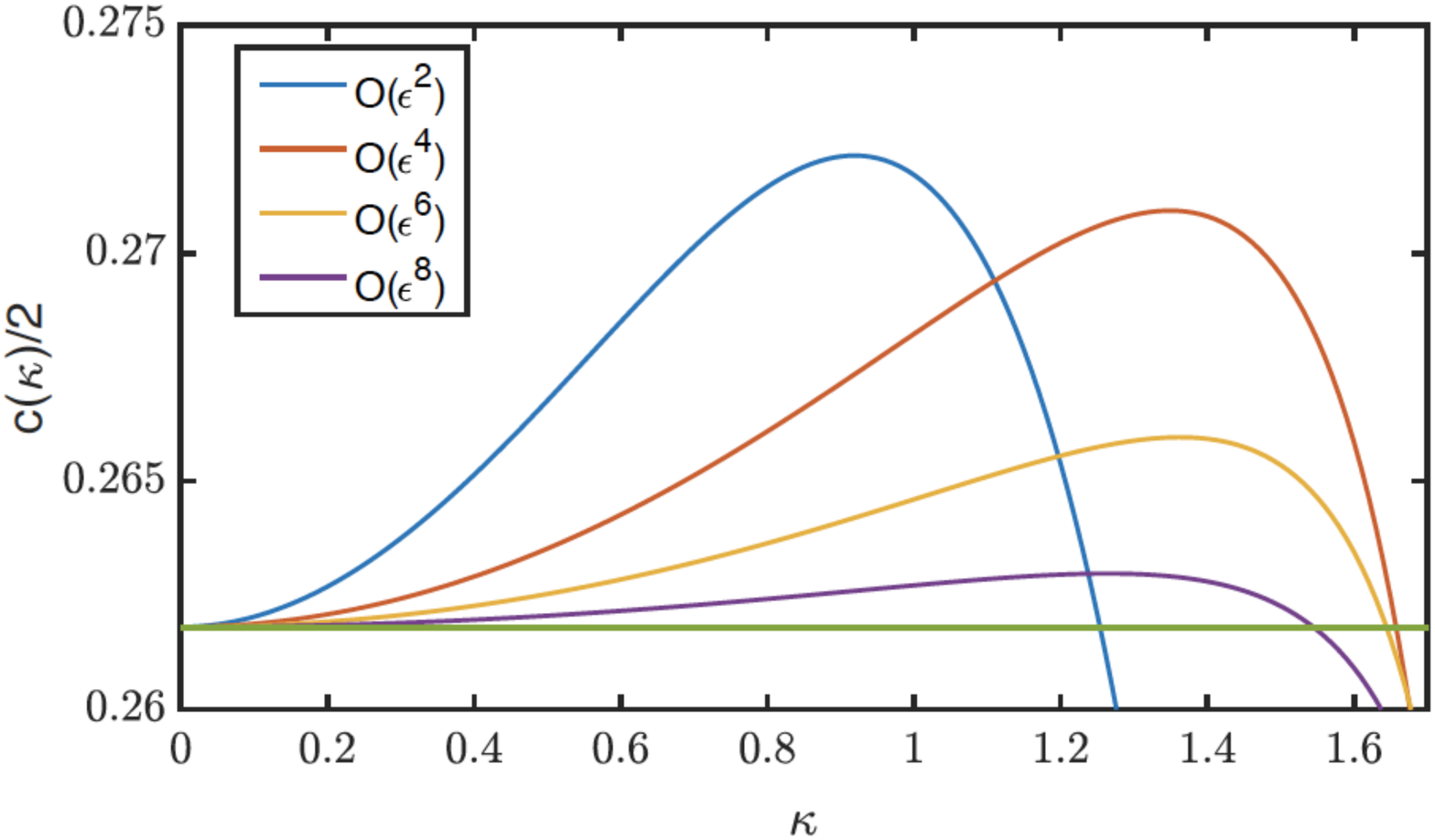}}
	\vspace{-8mm}
\caption{Specific heat coefficient $c/2$, from Eqs.(\ref{eq:trlog}), (\ref{eq:retarded_first}), (\ref{eq:free_en_expansion}), as a function of $\tilde{\kappa}$ for different orders in perturbation theory.}
\label{fig:kappa}
\end{figure}

One can observe in Fig.~\ref{fig:1} an order $10^{-3}$ correction in $c/2$ for small $\tilde{\kappa}$. While this correction is consistent with the exact diagonalisation results in Appendix A, it can also be an artefact of perturbation theory. We thus study the dependence of $c/2$ in $\tilde{\kappa}$ as we go higher orders in perturbation theory for $G^R(\epsilon)$. This is shown in Fig.~\ref{fig:kappa}.

The specific heat coefficient $c/2$ increases with $\tilde{\kappa}$. However, the higher order we go in perturbation theory for small frequencies $\epsilon$, the smaller is the increase for a given $\tilde{\kappa}$. This suggests that a fully non-perturbative calculation of $c(\tilde{\kappa})$ could lead to a $\tilde{\kappa}$ independent specific heat coefficient at least in the limit $\tilde{\kappa} \ll 1$.

Indeed this can also be understood analytically. First we need to identify which term gives this contribution. The integral over the $\pi/2$ factor in Eq.\eqref{eq:phaseshift} gives only a cutoff dependent contribution to the ground state energy $\frac{\beta\lambda}{2}$, and is unimportant in what concerns $c$. The remaining piece can again be studied using the Sommerfeld expansion,
\begin{align}
	-\beta\int_{-\lambda}^{\lambda}\frac{\dd\epsilon}{\pi}\frac{a(\epsilon,\tilde{\kappa},\beta)}{1+e^{\beta\epsilon}}\underset{\beta\gg 1}{=}&-\beta\int_{-\lambda}^{0}\frac{\dd\epsilon}{\pi}a(\epsilon,\tilde{\kappa},\beta)\notag
	\\&+\frac{\pi}{6\beta}\frac{\dd}{\dd\epsilon}\left.a(\epsilon,\tilde{\kappa},\beta)\right|_{\epsilon=0}+O(\beta^{-3}).
\end{align}
From Eq.\eqref{eq:retarded_first}, we can check that $\frac{\dd}{\dd\epsilon}\left.a\right|_{\epsilon=0} = -\frac{1}{2}$ for any $\tilde{\kappa}$ and $\beta$. Thus this term gives identical to that for $\tilde{\kappa}=0$. Since $a$ is now temperature dependent, possible corrections to the specific heat coefficient can come from the integral of $a$ over $(-\lambda,0)$ or from the higher order odd derivatives. However a simple series expansion of $a$ reveals that the leading order temperature dependence in $a$ comes at order $O(\epsilon^9)$, which is beyond the scope of the perturbative result in Eq.\eqref{eq:retarded_first}. One can check that, had we gone only to order $O(\epsilon^4)$ in $G^R$, the first order temperature dependence would have been at order $O(\epsilon^5)$. Going to order $O(\epsilon^{6})$ pushes the leading order temperature dependence of $a$ to order $O(\epsilon^7)$, and finally going to order $O(\epsilon^8)$ pushes it to $O(\epsilon^9)$. This analytical argument is fully consistent with the evaluation of $c$ from Eq.(\ref{eq:trlog}) depicted in Fig.\ref{fig:kappa}.

In Appendix A, the low temperature thermodynamic coefficients were studied numerically by exact diagonalisation. Corrections to the $\tilde{\kappa}=0$ specific heat coefficient $c =\pi/6$ and entropy density $s_0 = 0$ for $\tilde{\kappa}\ll 1$ were found to be of order $O(10^{-3})$ or lower. The analytic calculations from this Appendix confirm these results. They also corroborate the claim that the ground state of our model is dominated by the $\tilde{\kappa} = 0$ limit of Eq.(\ref{hami}). 

\section{Appendix C: Analytical calculation of the Lyapunov exponent with $q/2$-body interaction and $q\gg 1$}\label{appendix:C}
In this appendix we aim to give analytical support to the numerical results depicted in Fig. \ref{fig:Lyapunov} where it was shown that, for a fixed value of the coupling constant $\kappa > 0$, the Lyapunov exponent $\lambda_L$ always vanishes for sufficiently low temperatures. We have managed to obtain analytical results not for the Hamiltonian 
Eq.(\ref{hami}) but for a closely related model where the two-body term is replaced by a $q/2$-body, with $q\gg 1$, interaction keeping the one-body random perturbation of Eq.(\ref{hami}). The motivation for this choice is that, without the one-body perturbation, the Lyapunov exponent can be computed analytically at any coupling in the large $q$ limit, providing an explicit setup for studying the saturation of the chaos bound at low temperatures \cite{maldacena2016}.

By following closely the method of Ref. \cite{maldacena2016}, 
we show below that the perturbative expansion in $\kappa/J \ll 1$  captures the relevant physics. For any fixed $\kappa/J \ll 1$, we identify a range of temperatures for which the Lyapunov exponent is non-zero, though never saturates the bound on chaos. We also show that it vanishes for sufficiently low temperatures. This is in full agreement with the numerical results for the two-body model which corroborates the existence of a chaotic-to-integrable transition in this type of generalised SYK models.

{\it The model and the perturbative solution.-} 
As discussed in the main body of the Letter, we work with following Hamiltonian
\begin{align}\label{hami:large1}
\hspace{-1mm}H = i^{q\over 2} \hspace{-4mm}\sum_{1\leq i_1<i_2<\dots<i_q\leq N}\hspace{-4mm} J_{i_1, i_2,\dots, i_q}\ \chi_{i_1}\chi_{i_2} \dots \chi_{i_{q}} 
+i\sum_{1\leq i<j\leq N} \kappa_{ij} \  \chi_i \chi_j \,,
\end{align}
\noindent which is the $q/2$-interaction generalization of
the original model Eq.\eqref{eq:SYK4_hami}. As before $J_{i_1 i_2\dots i_q}$ and $\kappa_{ij}$ are Gaussian distributed random variables with zero average and variances $\frac{2^{q-1}}{q}\frac{(q-1)! J^2}{N^{q-1}}$ and $\frac{\kappa^2}{q N}$ respectively.
Note that, following \cite{maldacena2016}, we have rescaled the coupling constants: $\kappa^2\to \kappa^2/q$ and $J^2\to J^2 2^{q-1}/q$. The large $q$ limit is then defined by taking $q\gg 1$ and keeping the rescaled couplings fixed \cite{maldacena2016}. We  follow the same replica procedure described in the main body of the Letter to get the following effective action 
\begin{align}
S_\mathrm{eff}&=-\frac{1}{2}\Tr \log( \partial_\tau-\Sigma)+{1\over2}\int \dd \tau\dd\tau'\left[G(\tau,\tau')\Sigma(\tau,\tau')\right. \nonumber\\
&\left.-{ J^2\over 2^{1-q}q^2}G(\tau,\tau')^q-{\kappa^2\over 2q}G(\tau,\tau')^2\right].
\end{align}
As in \cite{maldacena2016}, in the limit $q\gg 1$ we can consistently expand $G$ as
\begin{align}
\label{largeq:expansion}
G(\tau) \underset{q\gg 1}{=} \frac{1}{2}\sgn(\tau)\left(1+\frac{1}{q} g(\tau)+O(q^{-2})\right).	
\end{align}
Inserting the above in the saddle point Eq.\eqref{eq:SP_fourier} and expanding in $q$, in Euclidean time we can simplify it to
\begin{align}
\label{largeq:eom}
\partial_{\theta}^{2}g = 2 (\beta  J)^2e^{g(\theta)} +(\beta \kappa)^2,
\end{align}
\noindent where $\theta = \tau/\beta \in [0,1)$. Together with the finite temperature boundary conditions $g(0)=g(1)=0$, this equation defines a non-linear boundary value problem for $g$. For $\kappa=0$, the solution is given by
\begin{align}
\label{zeroorder}
e^{g_{(0)}(\theta)} = \left[\frac{\cos{\frac{\pi\nu}{2}}}{\cos\left[\pi\nu\left(\frac{1}{2}-\theta\right)\right]}\right]^2,	&&\beta  J = \frac{\pi\nu}{\cos{\frac{\pi\nu}{2}}}.
\end{align}
Note that for $\nu=0$ we have $\beta J=0$ while for $\nu=1$, $\beta J=\infty$. Thus $\nu\in [0,1]$ parametrises the flow of $\beta J$. We  linearise Eq.\eqref{largeq:eom} around $g_{(0)}$ (the $\kappa=0$ solution). More explicitly, we substitute $g(\theta) = g_{(0)}(\theta)+\left(\frac{\kappa}{J}\right)^2 ~g_{(1)}(\theta)+O((\beta \kappa)^4)$ into Eq.\eqref{largeq:expansion} and then into Eq.\eqref{largeq:eom} to get the equation satisfied by $g_{(1)}$:  
\begin{align}\label{eq:g1}
\left(\partial_{x}^2 - \frac{2}{\cos^2(x)} \right)g_{(1)}(x) =\left(\frac{\beta J}{\pi\nu}\right)^2\,,
\end{align}	
\noindent where we changed  the $\theta$-coordinate to $x=\pi\nu\left(\frac{1}{2}-\theta\right)\in \left[-\frac{\pi\nu}{2},\frac{\pi\nu}{2}\right]$ and the corresponding boundary conditions are $g_{(1)}\left(\frac{\pi\nu}{2}\right)=g_{(1)}\left(-\frac{\pi\nu}{2}\right)=0$. The solution of this boundary-value problem is given by
\begin{align}
g_{(1)}(x)= \left(\frac{\beta J}{\pi\nu}\right)^2
&\Big[\alpha(x)\tan{x}+\log{\cos{x}}+x\tan{x}\notag\\
&+B(\nu)(x\tan{x}+1) \Big]\label{largeq:sol}
\end{align}
\noindent where we have defined,
\begin{widetext}
\begin{align}
\alpha(x) &= \int^x\dd t~\log\cos{t}=\frac{i}{2}  \text{Li}_2\left(-e^{2 i x}\right)+\frac{i x^2}{2}-x \log \left(1+e^{2 i
   x}\right)+x \log \cos (x)\\
B(\nu) &=-\frac{-\alpha \left(-\frac{\pi  \nu }{2}\right) \tan
   \left(\frac{\pi  \nu }{2}\right)+\alpha \left(\frac{\pi  \nu
   }{2}\right) \tan \left(\frac{\pi  \nu }{2}\right)+\pi  \nu 
   \tan \left(\frac{\pi  \nu }{2}\right)+2 \log \cos
   \left(\frac{\pi  \nu }{2}\right)}{\pi  \nu  \tan
   \left(\frac{\pi  \nu }{2}\right)+2}
\end{align}
\end{widetext}
Note that $B(\nu)$ is a negative monotonically decreasing function of $\nu$ bounded by $B(0) = 0$ and $B(1) = -1+\log{2}$.

{\it Lyapunov Exponent.-} As discussed in the main manuscript, the out-of-time -order four point correlator is generated by the convolution with the real-time retarded kernel Eq.\eqref{eq:otos}. At large $q$, this simplifies considerably,
\begin{align}
K_R(t_1,t_2,t_3,t_4)=\theta(t_{13})\theta(t_{24})\left[2 J^2 e^{g(\tau=it_{34}+\beta/2)}+q^{-1}\kappa^2 \right].
\end{align}
Thus, at large $q$ the second term is sub leading. By noting that $\partial_t\theta(t)=\delta(t)$, we can convert Eq.\eqref{eq:otos} from an integral equation to a differential equation by differentiating both sides with respect to $\partial_{t_{1}}$ and $\partial_{t_{2}}$,
\begin{align}
\partial_{t_1}\partial_{t_2}F(t_1,t_2) = 2 J^2 e^{g(\tau=it_{12}+\beta/2)}F(t_1,t_2)
\end{align}
Searching for solutions with exponential growth, $F(t_1,t_2) = e^{\frac{\lambda_L (t_1+t_2)}{2}}f(t_{12})$ and changing coordinates to $y=\frac{\pi\nu}{\beta} t_{12}$, the above simplifies to
\begin{align}
\label{eigen}
\left[\partial_{y}^2 +2\left(\frac{\beta J}{\pi\nu}\right)^2 e^{g(x\to iy)}\right] f(y) =\left(\frac{\beta\lambda_L}{2\pi\nu}\right)^2 f(y)
\end{align}
The equation above has the form of a one-dimensional Schr{\"o}dinger equation for the eigenfunction $f$ with eigenvalues $E_{\lambda} = -\left(\frac{\beta\lambda_L}{2\pi\nu}\right)^2$ in a potential $V(y) = -2\left(\frac{\beta J}{\pi\nu}\right)^2e^{g(iy)}$. For $\kappa=0$, $V_{(0)}(y) = 2\sech^2{y}$ is the well studied P{\"o}schl-Teller potential.  In particular, it has a bound state with energy $E^{(0)}_\lambda=-1$ and normalised eigenstate $f^{(0)}(y)= \frac{1}{\sqrt{2}}\sech{y}$. This implies that for this eigenstate we have $\lambda_L = \frac{2\pi}{\beta}\nu$ and therefore since $F(t_1,t_2) = e^{\lambda_L (t_1+t_2)/2}f(t_1-t_2)$ this bound state correspond to an exponential growth for the out-of -time-order four-point function.

\begin{figure}[h!]
\centering
	\hspace{-2mm}
\includegraphics[scale=0.47]{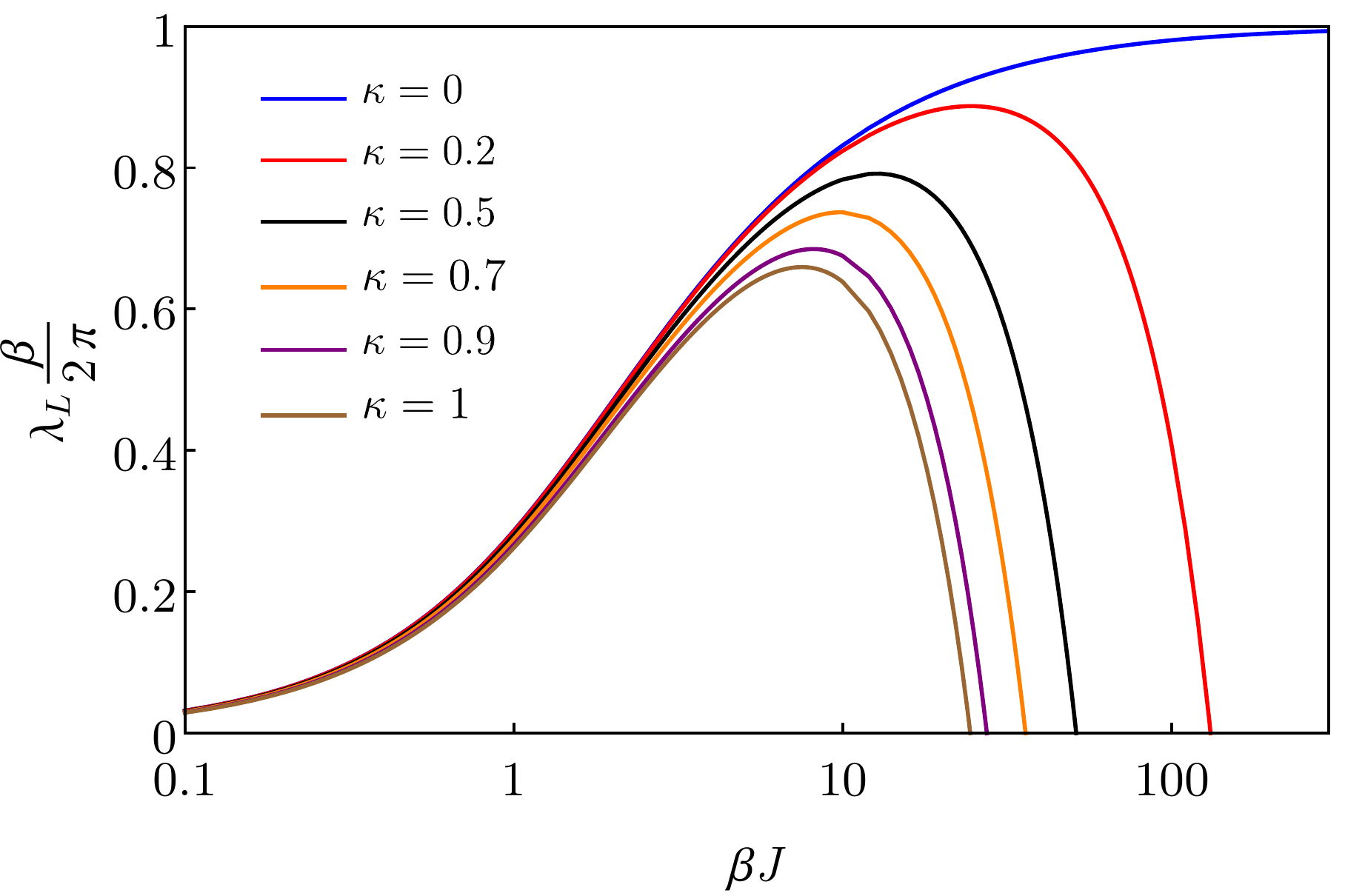}
	\vspace{-5mm}
\caption{Lyapunov exponent $\lambda_L$, Eq.(\ref{lyap}), as a function of the inverse temperature $\beta J$, in units of $1/k_B$, for different values of $\kappa$ and $ J=1$. Note that $\frac{\beta\lambda_L}{2\pi}\sim |\beta-\beta_*|$ vanishes linearly near the transition, with a slope of approximately $\frac{-2\kappa }{\pi\sqrt{72}}$. These results are in good quantitative agreement with those shown in Fig. \ref{fig:Lyapunov}.}
\label{fig:Lyapunov_ana}
\end{figure}

We are interested in studying what happens to this bound state when we add the $\kappa/J\ll 1$ correction to the potential. Or in a more suggestive notation, when we consider $V(y) = V_{(0)}(y)+\left(\frac{\kappa}{J}\right)^2 V_{(1)}(y)$ where $V_{(1)}(y) = e^{g_{(0)}(x\to iy)}g_{(1)}(x\to iy)$ with  $g_{(1)}(x)$ given by Eq.\eqref{largeq:sol}. By standard quantum mechanical perturbation theory, the correction $E_{\lambda} = E^{(0)}_{\lambda} + \left(\frac{\kappa}{J}\right)^2 E^{(1)}_{\lambda}$ is given by
\begin{widetext}
\begin{align}
E^{(1)}_{\lambda}& = \langle f^{(0)}|{2\over \cosh^2(y)} g_{(1)}(iy)|f^{(0)}\rangle =\half \int\limits_{-\infty}^{\infty} \dd y {g_{(1)}(iy)\over \cosh^4(y)} =
 {1\over 2}\left(\frac{\beta J}{\pi\nu}\right)^2\left[B(\nu)+{19\over 18}-\log2\right] = \left(\frac{\beta J}{\pi\nu}\right)^2\delta E (\nu)\,,
\end{align}
\end{widetext}
\noindent where $2\delta E (\nu) \equiv B(\nu)+{19\over 18}-\log2$. Therefore the correction to the bound-state energy is given by
\begin{align}
\label{energies}
E_{\lambda} = -1+\left(\frac{\beta \kappa}{\pi\nu}\right)^2\delta E(\nu)=-1+\left(\frac{ \kappa}{ J}\right)^2\frac{\delta E(\nu)}{\cos^2{\frac{\pi\nu}{2}}}.
\end{align}
Letting $E_{\lambda} = -\left(\frac{\beta\lambda}{2\pi\nu}\right)^2$ we obtain the correction to the Lyapunov exponent,
\begin{align}
	\label{lyap}
	\frac{\beta\lambda_L}{2\pi}=\sqrt{1-\left({\kappa\over J}\right)^2 {\delta E(\nu)\over \cos^2{\pi\nu\over2} }}\simeq 
	\nu-\left({\kappa\over J}\right)^2 {\nu \delta E(\nu)\over 2\cos^2{\pi\nu\over2} }+O\left(\left({\kappa\over J}\right)^4\right)\,.
\end{align}
In Fig.\ref{fig:Lyapunov_ana} we plot Lyapunov exponent for different values of $\kappa$ with $ J=1$.
This is to be compared with the exact numerical data from Fig.\eqref{fig:Lyapunov}. Although we do not get exact agreement between the critical values of $\beta_*$ for which $\lambda_L = 0$, the qualitative behaviour is similar. For $\kappa=0$ we get saturation of the bound at low temperatures. For any finite $\kappa>0$, there is a range of temperatures where a non-zero Lyapunov exponent persists, although this range decrease rapidly as we increase $\kappa$.

We estimate the critical temperature $\beta_*$ for which $\lambda_L$ crosses the real axis by doing a low temperature expansion of Eq.\eqref{lyap}. For $\beta J\gg 1$, we have
\begin{align}
1-\nu \underset{\beta J\gg 1}{=} \frac{2}{\beta J}-\frac{4}{(\beta J)^2}+\frac{24+\pi^2}{3(\beta J)^3}+O\left(\beta J^{-4}\right).
\end{align}
Inserting in Eq.\eqref{lyap} and expanding,
\begin{align}\label{eq:Lyap_lowT}
\frac{\beta\lambda}{2\pi} &=1-\frac{(\beta\kappa)^2}{\pi^2}\left[\frac{1}{72}+\frac{19-18\log \pi}{36\beta J}+O\left( {1\over (\beta J)^{2}}\right)\right].
\end{align}
Thus, assuming the transition occurs for large $\beta J$ we obtain that, to lowest order in $\beta \kappa$, the transition should occur when
\begin{align}
(\beta \kappa)_* ={\sqrt{72}\pi}.
\end{align}
This estimate gives $\beta_* \approx 133$ for $\kappa=0.2$ and $\beta_* \approx 53$ for $\kappa=0.5$, which is in very good agreement  with Fig.\ref{fig:Lyapunov_ana} and with the large-$N$ result obtained numerically for $q=4$ in Fig. \ref{fig:Lyapunov}. 

Similarly, one can also expand Eq.\eqref{lyap} in $\beta J\ll 1$. This gives the following high-temperature behaviour 
\begin{align}\label{eq:Lyap_largeT}
\frac{\beta\lambda}{2\pi}\underset{\beta J\ll 1}{=}&\frac{\beta J}{\pi}-\frac{(\beta J)^3}{8\pi}+\frac{(\beta \kappa)^2}{\pi}\left[\frac{-19+18\log{2}}{ \beta J}+O(\beta J)\right]\,,
\end{align}
where the ${\cal O}(\beta \kappa^2)$ correction is negative, indicating that chaos is weakened also in this regime of high temperatures.

\section{Appendix D: Anomalous value of $\langle r \rangle$ in the infrared limit of the spectrum}

\begin{figure}[H]
	\hspace{-5mm}\includegraphics[scale=1.12]{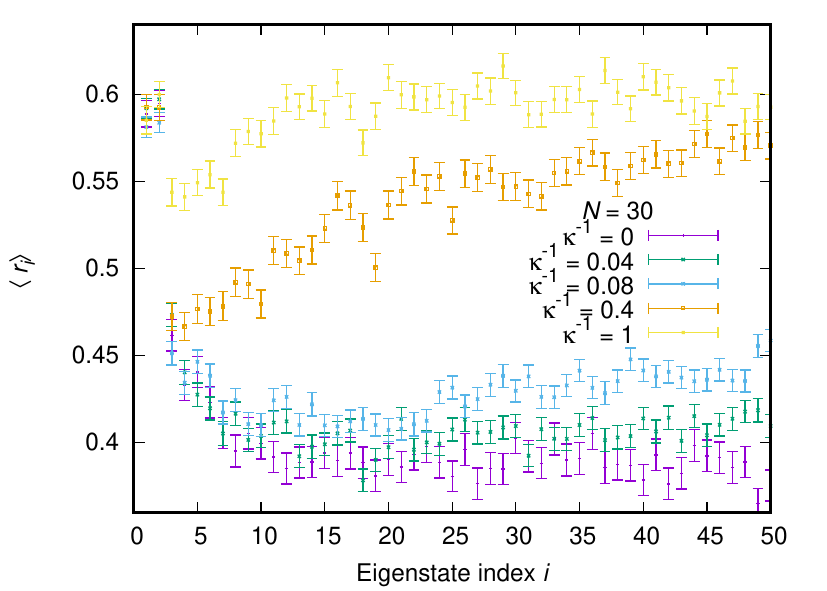}
	\vspace{-5mm}
	\caption{The sample average of the gap ratio $r_i = \min(\delta_i,\delta_{i+1})/\max(\delta_i,\delta_{i+1})$, in which $\delta_i=E_i-E_{i-1}$ is the difference between the neighboring energy eigenvalues in the ordered spectrum, plotted against $i$.}
	\label{fig:indexdep}
\end{figure}

In the main text of the Letter, we have studied the Sachdev-Ye-Kitaev model Eq. \eqref{hami} where we have taken $J=1$ as the unit of energy.
The eigenenergies of this model exhibit strong correlation at both ends of the spectrum
regardless of whether the model is modified with two-body interaction terms $\kappa>0$,
whereas it is less so if we are close to the center of the spectrum as the two-body terms get stronger.
In this supplemental material, we focus on the gap ratio 
\begin{align}
r_i = \frac{\min(\delta_i, \delta_{i+1})}{\max(\delta_i, \delta_{i+1})} 
\end{align}
as defined in the main text,
for an ordered spectrum $E_{i-1} < E_i < E_{i+1}$ where $\delta_i = E_i - E_{i-1}$.
In Figure~\ref{fig:indexdep} we have plotted the average gap ratio $\langle r_i\rangle$ against $i$
for several values of $\kappa$ and $N=30$.

We first observe that the values of $\langle r_i\rangle$ corresponding to the lowest 2 eigenvalues are very close to the value for the Gaussian unitary ensemble.
Also, the next $\sim 5-8$ eigenvalues are markedly larger (smaller) than the immediately following va\-lues if $\kappa$ is large (small).
After this, the dependence of $\langle r_i\rangle$ on $i$ is rather smooth but still significant for some choices of $\kappa$.
In the main text we removed $10$ eigenvalues from the lowest end of the spectrum for each sample.

Similar dependence of $\langle r_i\rangle$ on the eigenstate index $i$ is also observed for the other values of $N$.
In order to extract the features of the model that should survive for large $N$, we have removed ten lowest eigenvalues in obtaining $\langle r \rangle_\beta$
for the lower part of Fig.~2 in the main Letter.

\bibliographystyle{apsrev4-1}
\bibliography{library2}

\begin{thebibliography}{68}%
\makeatletter
\providecommand \@ifxundefined [1]{%
 \@ifx{#1\undefined}
}%
\providecommand \@ifnum [1]{%
 \ifnum #1\expandafter \@firstoftwo
 \else \expandafter \@secondoftwo
 \fi
}%
\providecommand \@ifx [1]{%
 \ifx #1\expandafter \@firstoftwo
 \else \expandafter \@secondoftwo
 \fi
}%
\providecommand \natexlab [1]{#1}%
\providecommand \enquote  [1]{``#1''}%
\providecommand \bibnamefont  [1]{#1}%
\providecommand \bibfnamefont [1]{#1}%
\providecommand \citenamefont [1]{#1}%
\providecommand \href@noop [0]{\@secondoftwo}%
\providecommand \href [0]{\begingroup \@sanitize@url \@href}%
\providecommand \@href[1]{\@@startlink{#1}\@@href}%
\providecommand \@@href[1]{\endgroup#1\@@endlink}%
\providecommand \@sanitize@url [0]{\catcode `\\12\catcode `\$12\catcode
  `\&12\catcode `\#12\catcode `\^12\catcode `\_12\catcode `\%12\relax}%
\providecommand \@@startlink[1]{}%
\providecommand \@@endlink[0]{}%
\providecommand \url  [0]{\begingroup\@sanitize@url \@url }%
\providecommand \@url [1]{\endgroup\@href {#1}{\urlprefix }}%
\providecommand \urlprefix  [0]{URL }%
\providecommand \Eprint [0]{\href }%
\providecommand \doibase [0]{http://dx.doi.org/}%
\providecommand \selectlanguage [0]{\@gobble}%
\providecommand \bibinfo  [0]{\@secondoftwo}%
\providecommand \bibfield  [0]{\@secondoftwo}%
\providecommand \translation [1]{[#1]}%
\providecommand \BibitemOpen [0]{}%
\providecommand \bibitemStop [0]{}%
\providecommand \bibitemNoStop [0]{.\EOS\space}%
\providecommand \EOS [0]{\spacefactor3000\relax}%
\providecommand \BibitemShut  [1]{\csname bibitem#1\endcsname}%
\let\auto@bib@innerbib\@empty
\bibitem [{\citenamefont {Bohigas}\ and\ \citenamefont
  {Flores}(1971{\natexlab{a}})}]{bohigas1971}%
  \BibitemOpen
  \bibfield  {author} {\bibinfo {author} {\bibfnamefont {O.}~\bibnamefont
  {Bohigas}}\ and\ \bibinfo {author} {\bibfnamefont {J.}~\bibnamefont
  {Flores}},\ }\href {\doibase http://dx.doi.org/10.1016/0370-2693(71)90598-3}
  {\bibfield  {journal} {\bibinfo  {journal} {Physics Letters B}\ }\textbf
  {\bibinfo {volume} {34}},\ \bibinfo {pages} {261 } (\bibinfo {year}
  {1971}{\natexlab{a}})}\BibitemShut {NoStop}%
\bibitem [{\citenamefont {Bohigas}\ and\ \citenamefont
  {Flores}(1971{\natexlab{b}})}]{bohigas1971a}%
  \BibitemOpen
  \bibfield  {author} {\bibinfo {author} {\bibfnamefont {O.}~\bibnamefont
  {Bohigas}}\ and\ \bibinfo {author} {\bibfnamefont {J.}~\bibnamefont
  {Flores}},\ }\href {\doibase http://dx.doi.org/10.1016/0370-2693(71)90399-6}
  {\bibfield  {journal} {\bibinfo  {journal} {Physics Letters B}\ }\textbf
  {\bibinfo {volume} {35}},\ \bibinfo {pages} {383 } (\bibinfo {year}
  {1971}{\natexlab{b}})}\BibitemShut {NoStop}%
\bibitem [{\citenamefont {French}\ and\ \citenamefont
  {Wong}(1970)}]{french1970}%
  \BibitemOpen
  \bibfield  {author} {\bibinfo {author} {\bibfnamefont {J.}~\bibnamefont
  {French}}\ and\ \bibinfo {author} {\bibfnamefont {S.}~\bibnamefont {Wong}},\
  }\href {\doibase http://dx.doi.org/10.1016/0370-2693(70)90213-3} {\bibfield
  {journal} {\bibinfo  {journal} {Physics Letters B}\ }\textbf {\bibinfo
  {volume} {33}},\ \bibinfo {pages} {449 } (\bibinfo {year}
  {1970})}\BibitemShut {NoStop}%
\bibitem [{\citenamefont {French}\ and\ \citenamefont
  {Wong}(1971)}]{french1971}%
  \BibitemOpen
  \bibfield  {author} {\bibinfo {author} {\bibfnamefont {J.}~\bibnamefont
  {French}}\ and\ \bibinfo {author} {\bibfnamefont {S.}~\bibnamefont {Wong}},\
  }\href {\doibase http://dx.doi.org/10.1016/0370-2693(71)90424-2} {\bibfield
  {journal} {\bibinfo  {journal} {Physics Letters B}\ }\textbf {\bibinfo
  {volume} {35}},\ \bibinfo {pages} {5 } (\bibinfo {year} {1971})}\BibitemShut
  {NoStop}%
\bibitem [{\citenamefont {Mon}\ and\ \citenamefont {French}(1975)}]{mon1975}%
  \BibitemOpen
  \bibfield  {author} {\bibinfo {author} {\bibfnamefont {K.}~\bibnamefont
  {Mon}}\ and\ \bibinfo {author} {\bibfnamefont {J.}~\bibnamefont {French}},\
  }\href {\doibase http://dx.doi.org/10.1016/0003-4916(75)90045-7} {\bibfield
  {journal} {\bibinfo  {journal} {Annals of Physics}\ }\textbf {\bibinfo
  {volume} {95}},\ \bibinfo {pages} {90 } (\bibinfo {year} {1975})}\BibitemShut
  {NoStop}%
\bibitem [{\citenamefont {Benet}\ and\ \citenamefont
  {Weidenm{\"u}ller}(2003)}]{benet2003}%
  \BibitemOpen
  \bibfield  {author} {\bibinfo {author} {\bibfnamefont {L.}~\bibnamefont
  {Benet}}\ and\ \bibinfo {author} {\bibfnamefont {H.~A.}\ \bibnamefont
  {Weidenm{\"u}ller}},\ }\href {http://stacks.iop.org/0305-4470/36/i=12/a=340}
  {\bibfield  {journal} {\bibinfo  {journal} {Journal of Physics A:
  Mathematical and General}\ }\textbf {\bibinfo {volume} {36}},\ \bibinfo
  {pages} {3569} (\bibinfo {year} {2003})}\BibitemShut {NoStop}%
\bibitem [{\citenamefont {Kota}(2014)}]{kota2014}%
  \BibitemOpen
  \bibfield  {author} {\bibinfo {author} {\bibfnamefont {V.~K.~B.}\
  \bibnamefont {Kota}},\ }\href@noop {} {\emph {\bibinfo {title} {Embedded
  random matrix ensembles in quantum physics}}},\ Vol.\ \bibinfo {volume}
  {884}\ (\bibinfo  {publisher} {Springer},\ \bibinfo {year}
  {2014})\BibitemShut {NoStop}%
\bibitem [{\citenamefont {Sachdev}\ and\ \citenamefont
  {Ye}(1993)}]{sachdev1993}%
  \BibitemOpen
  \bibfield  {author} {\bibinfo {author} {\bibfnamefont {S.}~\bibnamefont
  {Sachdev}}\ and\ \bibinfo {author} {\bibfnamefont {J.}~\bibnamefont {Ye}},\
  }\href {\doibase 10.1103/PhysRevLett.70.3339} {\bibfield  {journal} {\bibinfo
   {journal} {Phys. Rev. Lett.}\ }\textbf {\bibinfo {volume} {70}},\ \bibinfo
  {pages} {3339} (\bibinfo {year} {1993})}\BibitemShut {NoStop}%
\bibitem [{\citenamefont {Sachdev}(2010)}]{sachdev2010}%
  \BibitemOpen
  \bibfield  {author} {\bibinfo {author} {\bibfnamefont {S.}~\bibnamefont
  {Sachdev}},\ }\href {\doibase 10.1103/PhysRevLett.105.151602} {\bibfield
  {journal} {\bibinfo  {journal} {Phys. Rev. Lett.}\ }\textbf {\bibinfo
  {volume} {105}},\ \bibinfo {pages} {151602} (\bibinfo {year}
  {2010})}\BibitemShut {NoStop}%
\bibitem [{\citenamefont {Kitaev}()}]{kitaev2015}%
  \BibitemOpen
  \bibfield  {author} {\bibinfo {author} {\bibfnamefont {A.}~\bibnamefont
  {Kitaev}},\ }\href@noop {} {\enquote {\bibinfo {title} {A simple model of
  quantum holography},}\ }\bibinfo {note} {KITP strings seminar and
  Entanglement 2015 program, 12 February, 7 April and 27 May 2015,
  http://online.kitp.ucsb.edu/online/entangled15/}\BibitemShut {NoStop}%
\bibitem [{\citenamefont {Jensen}(2016)}]{jensen2016}%
  \BibitemOpen
  \bibfield  {author} {\bibinfo {author} {\bibfnamefont {K.}~\bibnamefont
  {Jensen}},\ }\href {\doibase 10.1103/PhysRevLett.117.111601} {\bibfield
  {journal} {\bibinfo  {journal} {Phys. Rev. Lett.}\ }\textbf {\bibinfo
  {volume} {117}},\ \bibinfo {pages} {111601} (\bibinfo {year}
  {2016})}\BibitemShut {NoStop}%
\bibitem [{\citenamefont {Maldacena}\ and\ \citenamefont
  {Stanford}(2016)}]{maldacena2016}%
  \BibitemOpen
  \bibfield  {author} {\bibinfo {author} {\bibfnamefont {J.}~\bibnamefont
  {Maldacena}}\ and\ \bibinfo {author} {\bibfnamefont {D.}~\bibnamefont
  {Stanford}},\ }\href {\doibase 10.1103/PhysRevD.94.106002} {\bibfield
  {journal} {\bibinfo  {journal} {Phys. Rev. D}\ }\textbf {\bibinfo {volume}
  {94}},\ \bibinfo {pages} {106002} (\bibinfo {year} {2016})}\BibitemShut
  {NoStop}%
\bibitem [{\citenamefont {Sachdev}(2015)}]{sachdev2015}%
  \BibitemOpen
  \bibfield  {author} {\bibinfo {author} {\bibfnamefont {S.}~\bibnamefont
  {Sachdev}},\ }\href {\doibase 10.1103/PhysRevX.5.041025} {\bibfield
  {journal} {\bibinfo  {journal} {Phys. Rev. X}\ }\textbf {\bibinfo {volume}
  {5}},\ \bibinfo {pages} {041025} (\bibinfo {year} {2015})}\BibitemShut
  {NoStop}%
\bibitem [{\citenamefont {Almheiri}\ and\ \citenamefont
  {Polchinski}(2015)}]{almheiri2015}%
  \BibitemOpen
  \bibfield  {author} {\bibinfo {author} {\bibfnamefont {A.}~\bibnamefont
  {Almheiri}}\ and\ \bibinfo {author} {\bibfnamefont {J.}~\bibnamefont
  {Polchinski}},\ }\href {\doibase 10.1007/JHEP11(2015)014} {\bibfield
  {journal} {\bibinfo  {journal} {Journal of High Energy Physics}\ }\textbf
  {\bibinfo {volume} {11}},\ \bibinfo {pages} {1} (\bibinfo {year}
  {2015})}\BibitemShut {NoStop}%
\bibitem [{\citenamefont {Maldacena}\ \emph
  {et~al.}(2016{\natexlab{a}})\citenamefont {Maldacena}, \citenamefont
  {Stanford},\ and\ \citenamefont {Yang}}]{maldacena2016a}%
  \BibitemOpen
  \bibfield  {author} {\bibinfo {author} {\bibfnamefont {J.}~\bibnamefont
  {Maldacena}}, \bibinfo {author} {\bibfnamefont {D.}~\bibnamefont {Stanford}},
  \ and\ \bibinfo {author} {\bibfnamefont {Z.}~\bibnamefont {Yang}},\ }\href
  {\doibase 10.1093/ptep/ptw124} {\bibfield  {journal} {\bibinfo  {journal}
  {Progress of Theoretical and Experimental Physics}\ }\textbf {\bibinfo
  {volume} {2016}},\ \bibinfo {pages} {12C104} (\bibinfo {year}
  {2016}{\natexlab{a}})}\BibitemShut {NoStop}%
\bibitem [{\citenamefont {Engels{\"o}y}\ \emph {et~al.}(2016)\citenamefont
  {Engels{\"o}y}, \citenamefont {Mertens},\ and\ \citenamefont
  {Verlinde}}]{engels2016}%
  \BibitemOpen
  \bibfield  {author} {\bibinfo {author} {\bibfnamefont {J.}~\bibnamefont
  {Engels{\"o}y}}, \bibinfo {author} {\bibfnamefont {T.~G.}\ \bibnamefont
  {Mertens}}, \ and\ \bibinfo {author} {\bibfnamefont {H.}~\bibnamefont
  {Verlinde}},\ }\href {\doibase 10.1007/JHEP07(2016)139} {\bibfield  {journal}
  {\bibinfo  {journal} {Journal of High Energy Physics}\ }\textbf {\bibinfo
  {volume} {07}},\ \bibinfo {pages} {1} (\bibinfo {year} {2016})}\BibitemShut
  {NoStop}%
\bibitem [{\citenamefont {Bagrets}\ \emph {et~al.}(2016)\citenamefont
  {Bagrets}, \citenamefont {Altland},\ and\ \citenamefont
  {Kamenev}}]{bagrets2016}%
  \BibitemOpen
  \bibfield  {author} {\bibinfo {author} {\bibfnamefont {D.}~\bibnamefont
  {Bagrets}}, \bibinfo {author} {\bibfnamefont {A.}~\bibnamefont {Altland}}, \
  and\ \bibinfo {author} {\bibfnamefont {A.}~\bibnamefont {Kamenev}},\
  }\href@noop {} {\bibfield  {journal} {\bibinfo  {journal} {Nuclear Physics
  B}\ }\textbf {\bibinfo {volume} {911}},\ \bibinfo {pages} {191} (\bibinfo
  {year} {2016})}\BibitemShut {NoStop}%
\bibitem [{\citenamefont {Jian}\ \emph {et~al.}(2017)\citenamefont {Jian},
  \citenamefont {Bi},\ and\ \citenamefont {Xu}}]{cenke2017}%
  \BibitemOpen
  \bibfield  {author} {\bibinfo {author} {\bibfnamefont {C.-M.}\ \bibnamefont
  {Jian}}, \bibinfo {author} {\bibfnamefont {Z.}~\bibnamefont {Bi}}, \ and\
  \bibinfo {author} {\bibfnamefont {C.}~\bibnamefont {Xu}},\ }\href {\doibase
  10.1103/PhysRevB.96.115122} {\bibfield  {journal} {\bibinfo  {journal} {Phys.
  Rev. B}\ }\textbf {\bibinfo {volume} {96}},\ \bibinfo {pages} {115122}
  (\bibinfo {year} {2017})}\BibitemShut {NoStop}%
\bibitem [{\citenamefont {Danshita}\ \emph {et~al.}(2017)\citenamefont
  {Danshita}, \citenamefont {Hanada},\ and\ \citenamefont
  {Tezuka}}]{danshita2016}%
  \BibitemOpen
  \bibfield  {author} {\bibinfo {author} {\bibfnamefont {I.}~\bibnamefont
  {Danshita}}, \bibinfo {author} {\bibfnamefont {M.}~\bibnamefont {Hanada}}, \
  and\ \bibinfo {author} {\bibfnamefont {M.}~\bibnamefont {Tezuka}},\
  }\href@noop {} {\bibfield  {journal} {\bibinfo  {journal} {Prog. Theor. Exp.
  Phys.}\ }\textbf {\bibinfo {volume} {2017}},\ \bibinfo {pages} {083I01}
  (\bibinfo {year} {2017})}\BibitemShut {NoStop}%
\bibitem [{\citenamefont {Jevicki}\ \emph {et~al.}(2016)\citenamefont
  {Jevicki}, \citenamefont {Suzuki},\ and\ \citenamefont {Yoon}}]{jevicki2016}%
  \BibitemOpen
  \bibfield  {author} {\bibinfo {author} {\bibfnamefont {A.}~\bibnamefont
  {Jevicki}}, \bibinfo {author} {\bibfnamefont {K.}~\bibnamefont {Suzuki}}, \
  and\ \bibinfo {author} {\bibfnamefont {J.}~\bibnamefont {Yoon}},\ }\href
  {\doibase 10.1007/JHEP07(2016)007} {\bibfield  {journal} {\bibinfo  {journal}
  {Journal of High Energy Physics}\ }\textbf {\bibinfo {volume} {07}},\
  \bibinfo {pages} {1} (\bibinfo {year} {2016})}\BibitemShut {NoStop}%
\bibitem [{\citenamefont {Jian}\ and\ \citenamefont {Yao}(2017)}]{jian2017}%
  \BibitemOpen
  \bibfield  {author} {\bibinfo {author} {\bibfnamefont {S.-K.}\ \bibnamefont
  {Jian}}\ and\ \bibinfo {author} {\bibfnamefont {H.}~\bibnamefont {Yao}},\
  }\href {\doibase 10.1103/PhysRevLett.119.206602} {\bibfield  {journal}
  {\bibinfo  {journal} {Phys. Rev. Lett.}\ }\textbf {\bibinfo {volume} {119}},\
  \bibinfo {pages} {206602} (\bibinfo {year} {2017})}\BibitemShut {NoStop}%
\bibitem [{\citenamefont {Mag\'an}(2016)}]{magan2016}%
  \BibitemOpen
  \bibfield  {author} {\bibinfo {author} {\bibfnamefont {J.~M.}\ \bibnamefont
  {Mag\'an}},\ }\href {\doibase 10.1103/PhysRevLett.116.030401} {\bibfield
  {journal} {\bibinfo  {journal} {Phys. Rev. Lett.}\ }\textbf {\bibinfo
  {volume} {116}},\ \bibinfo {pages} {030401} (\bibinfo {year}
  {2016})}\BibitemShut {NoStop}%
\bibitem [{\citenamefont {Mag{\'a}n}(2016)}]{magan2016b}%
  \BibitemOpen
  \bibfield  {author} {\bibinfo {author} {\bibfnamefont {J.~M.}\ \bibnamefont
  {Mag{\'a}n}},\ }\href {\doibase 10.1007/JHEP08(2016)081} {\bibfield
  {journal} {\bibinfo  {journal} {Journal of High Energy Physics}\ }\textbf
  {\bibinfo {volume} {2016}},\ \bibinfo {pages} {81} (\bibinfo {year}
  {2016})}\BibitemShut {NoStop}%
\bibitem [{\citenamefont {Witten}(2016)}]{witten2016}%
  \BibitemOpen
  \bibfield  {author} {\bibinfo {author} {\bibfnamefont {E.}~\bibnamefont
  {Witten}},\ }\href@noop {} {\  (\bibinfo {year} {2016})},\ \Eprint
  {http://arxiv.org/abs/1610.09758} {arXiv:1610.09758 [hep-th]} \BibitemShut
  {NoStop}%
\bibitem [{\citenamefont {You}\ \emph {et~al.}(2017)\citenamefont {You},
  \citenamefont {Ludwig},\ and\ \citenamefont {Xu}}]{you2016}%
  \BibitemOpen
  \bibfield  {author} {\bibinfo {author} {\bibfnamefont {Y.-Z.}\ \bibnamefont
  {You}}, \bibinfo {author} {\bibfnamefont {A.~W.~W.}\ \bibnamefont {Ludwig}},
  \ and\ \bibinfo {author} {\bibfnamefont {C.}~\bibnamefont {Xu}},\ }\href
  {\doibase 10.1103/PhysRevB.95.115150} {\bibfield  {journal} {\bibinfo
  {journal} {Phys. Rev. B}\ }\textbf {\bibinfo {volume} {95}},\ \bibinfo
  {pages} {115150} (\bibinfo {year} {2017})}\BibitemShut {NoStop}%
\bibitem [{\citenamefont {Garc\'{\i}a-Garc\'{\i}a}\ and\ \citenamefont
  {Verbaarschot}(2016)}]{garcia2016}%
  \BibitemOpen
  \bibfield  {author} {\bibinfo {author} {\bibfnamefont {A.~M.}\ \bibnamefont
  {Garc\'{\i}a-Garc\'{\i}a}}\ and\ \bibinfo {author} {\bibfnamefont {J.~J.~M.}\
  \bibnamefont {Verbaarschot}},\ }\href {\doibase 10.1103/PhysRevD.94.126010}
  {\bibfield  {journal} {\bibinfo  {journal} {Phys. Rev. D}\ }\textbf {\bibinfo
  {volume} {94}},\ \bibinfo {pages} {126010} (\bibinfo {year}
  {2016})}\BibitemShut {NoStop}%
\bibitem [{\citenamefont {Garc\'{\i}a-Garc\'{\i}a}\ and\ \citenamefont
  {Verbaarschot}(2017)}]{garcia2017}%
  \BibitemOpen
  \bibfield  {author} {\bibinfo {author} {\bibfnamefont {A.~M.}\ \bibnamefont
  {Garc\'{\i}a-Garc\'{\i}a}}\ and\ \bibinfo {author} {\bibfnamefont {J.~J.~M.}\
  \bibnamefont {Verbaarschot}},\ }\href {\doibase 10.1103/PhysRevD.96.066012}
  {\bibfield  {journal} {\bibinfo  {journal} {Phys. Rev. D}\ }\textbf {\bibinfo
  {volume} {96}},\ \bibinfo {pages} {066012} (\bibinfo {year}
  {2017})}\BibitemShut {NoStop}%
\bibitem [{\citenamefont {Klebanov}\ and\ \citenamefont
  {Tarnopolsky}(2017)}]{klebanov2017}%
  \BibitemOpen
  \bibfield  {author} {\bibinfo {author} {\bibfnamefont {I.~R.}\ \bibnamefont
  {Klebanov}}\ and\ \bibinfo {author} {\bibfnamefont {G.}~\bibnamefont
  {Tarnopolsky}},\ }\href {\doibase 10.1103/PhysRevD.95.046004} {\bibfield
  {journal} {\bibinfo  {journal} {Phys. Rev. D}\ }\textbf {\bibinfo {volume}
  {95}},\ \bibinfo {pages} {046004} (\bibinfo {year} {2017})}\BibitemShut
  {NoStop}%
\bibitem [{\citenamefont {Bagrets}\ \emph {et~al.}(2017)\citenamefont
  {Bagrets}, \citenamefont {Altland},\ and\ \citenamefont
  {Kamenev}}]{bagrets2017}%
  \BibitemOpen
  \bibfield  {author} {\bibinfo {author} {\bibfnamefont {D.}~\bibnamefont
  {Bagrets}}, \bibinfo {author} {\bibfnamefont {A.}~\bibnamefont {Altland}}, \
  and\ \bibinfo {author} {\bibfnamefont {A.}~\bibnamefont {Kamenev}},\ }\href
  {\doibase https://doi.org/10.1016/j.nuclphysb.2017.06.012} {\bibfield
  {journal} {\bibinfo  {journal} {Nuclear Physics B}\ }\textbf {\bibinfo
  {volume} {921}},\ \bibinfo {pages} {727 } (\bibinfo {year}
  {2017})}\BibitemShut {NoStop}%
\bibitem [{\citenamefont {Cotler}\ \emph {et~al.}(2017)\citenamefont {Cotler},
  \citenamefont {Gur-Ari}, \citenamefont {Hanada}, \citenamefont {Polchinski},
  \citenamefont {Saad}, \citenamefont {Shenker}, \citenamefont {Stanford},
  \citenamefont {Streicher},\ and\ \citenamefont {Tezuka}}]{cotler2016}%
  \BibitemOpen
  \bibfield  {author} {\bibinfo {author} {\bibfnamefont {J.~S.}\ \bibnamefont
  {Cotler}}, \bibinfo {author} {\bibfnamefont {G.}~\bibnamefont {Gur-Ari}},
  \bibinfo {author} {\bibfnamefont {M.}~\bibnamefont {Hanada}}, \bibinfo
  {author} {\bibfnamefont {J.}~\bibnamefont {Polchinski}}, \bibinfo {author}
  {\bibfnamefont {P.}~\bibnamefont {Saad}}, \bibinfo {author} {\bibfnamefont
  {S.~H.}\ \bibnamefont {Shenker}}, \bibinfo {author} {\bibfnamefont
  {D.}~\bibnamefont {Stanford}}, \bibinfo {author} {\bibfnamefont
  {A.}~\bibnamefont {Streicher}}, \ and\ \bibinfo {author} {\bibfnamefont
  {M.}~\bibnamefont {Tezuka}},\ }\href {\doibase 10.1007/JHEP05(2017)118}
  {\bibfield  {journal} {\bibinfo  {journal} {Journal of High Energy Physics}\
  }\textbf {\bibinfo {volume} {05}},\ \bibinfo {pages} {118} (\bibinfo {year}
  {2017})}\BibitemShut {NoStop}%
\bibitem [{\citenamefont {Banerjee}\ and\ \citenamefont
  {Altman}(2017)}]{altman2017}%
  \BibitemOpen
  \bibfield  {author} {\bibinfo {author} {\bibfnamefont {S.}~\bibnamefont
  {Banerjee}}\ and\ \bibinfo {author} {\bibfnamefont {E.}~\bibnamefont
  {Altman}},\ }\href {\doibase 10.1103/PhysRevB.95.134302} {\bibfield
  {journal} {\bibinfo  {journal} {Phys. Rev. B}\ }\textbf {\bibinfo {volume}
  {95}},\ \bibinfo {pages} {134302} (\bibinfo {year} {2017})}\BibitemShut
  {NoStop}%
\bibitem [{\citenamefont {Kanazawa}\ and\ \citenamefont
  {Wettig}(2017)}]{kanazawa2017}%
  \BibitemOpen
  \bibfield  {author} {\bibinfo {author} {\bibfnamefont {T.}~\bibnamefont
  {Kanazawa}}\ and\ \bibinfo {author} {\bibfnamefont {T.}~\bibnamefont
  {Wettig}},\ }\href {\doibase 10.1007/JHEP09(2017)050} {\bibfield  {journal}
  {\bibinfo  {journal} {Journal of High Energy Physics}\ }\textbf {\bibinfo
  {volume} {2017}},\ \bibinfo {pages} {50} (\bibinfo {year}
  {2017})}\BibitemShut {NoStop}%
\bibitem [{\citenamefont {Krishnan}\ \emph
  {et~al.}(2017{\natexlab{a}})\citenamefont {Krishnan}, \citenamefont
  {Sanyal},\ and\ \citenamefont {Subramanian}}]{krishnan2017}%
  \BibitemOpen
  \bibfield  {author} {\bibinfo {author} {\bibfnamefont {C.}~\bibnamefont
  {Krishnan}}, \bibinfo {author} {\bibfnamefont {S.}~\bibnamefont {Sanyal}}, \
  and\ \bibinfo {author} {\bibfnamefont {P.~N.~B.}\ \bibnamefont
  {Subramanian}},\ }\href {\doibase 10.1007/JHEP03(2017)056} {\bibfield
  {journal} {\bibinfo  {journal} {Journal of High Energy Physics}\ }\textbf
  {\bibinfo {volume} {2017}},\ \bibinfo {pages} {56} (\bibinfo {year}
  {2017}{\natexlab{a}})}\BibitemShut {NoStop}%
\bibitem [{\citenamefont {Krishnan}\ \emph
  {et~al.}(2017{\natexlab{b}})\citenamefont {Krishnan}, \citenamefont {Kumar},\
  and\ \citenamefont {Sanyal}}]{krishnan2017a}%
  \BibitemOpen
  \bibfield  {author} {\bibinfo {author} {\bibfnamefont {C.}~\bibnamefont
  {Krishnan}}, \bibinfo {author} {\bibfnamefont {K.~V.~P.}\ \bibnamefont
  {Kumar}}, \ and\ \bibinfo {author} {\bibfnamefont {S.}~\bibnamefont
  {Sanyal}},\ }\href {\doibase 10.1007/JHEP06(2017)036} {\bibfield  {journal}
  {\bibinfo  {journal} {Journal of High Energy Physics}\ }\textbf {\bibinfo
  {volume} {2017}},\ \bibinfo {pages} {1} (\bibinfo {year}
  {2017}{\natexlab{b}})}\BibitemShut {NoStop}%
\bibitem [{\citenamefont {Garc\'{\i}a-\'Alvarez}\ \emph
  {et~al.}(2017)\citenamefont {Garc\'{\i}a-\'Alvarez}, \citenamefont
  {Egusquiza}, \citenamefont {Lamata}, \citenamefont {del Campo}, \citenamefont
  {Sonner},\ and\ \citenamefont {Solano}}]{garcia-alvarez2016}%
  \BibitemOpen
  \bibfield  {author} {\bibinfo {author} {\bibfnamefont {L.}~\bibnamefont
  {Garc\'{\i}a-\'Alvarez}}, \bibinfo {author} {\bibfnamefont {I.~L.}\
  \bibnamefont {Egusquiza}}, \bibinfo {author} {\bibfnamefont {L.}~\bibnamefont
  {Lamata}}, \bibinfo {author} {\bibfnamefont {A.}~\bibnamefont {del Campo}},
  \bibinfo {author} {\bibfnamefont {J.}~\bibnamefont {Sonner}}, \ and\ \bibinfo
  {author} {\bibfnamefont {E.}~\bibnamefont {Solano}},\ }\href {\doibase
  10.1103/PhysRevLett.119.040501} {\bibfield  {journal} {\bibinfo  {journal}
  {Phys. Rev. Lett.}\ }\textbf {\bibinfo {volume} {119}},\ \bibinfo {pages}
  {040501} (\bibinfo {year} {2017})}\BibitemShut {NoStop}%
\bibitem [{\citenamefont {Pikulin}\ and\ \citenamefont
  {Franz}(2017)}]{Pikulin2017}%
  \BibitemOpen
  \bibfield  {author} {\bibinfo {author} {\bibfnamefont {D.~I.}\ \bibnamefont
  {Pikulin}}\ and\ \bibinfo {author} {\bibfnamefont {M.}~\bibnamefont
  {Franz}},\ }\href {\doibase 10.1103/PhysRevX.7.031006} {\bibfield  {journal}
  {\bibinfo  {journal} {Phys. Rev. X}\ }\textbf {\bibinfo {volume} {7}},\
  \bibinfo {pages} {031006} (\bibinfo {year} {2017})}\BibitemShut {NoStop}%
\bibitem [{\citenamefont {Gu}\ \emph {et~al.}(2017)\citenamefont {Gu},
  \citenamefont {Qi},\ and\ \citenamefont {Stanford}}]{gu2016}%
  \BibitemOpen
  \bibfield  {author} {\bibinfo {author} {\bibfnamefont {Y.}~\bibnamefont
  {Gu}}, \bibinfo {author} {\bibfnamefont {X.-L.}\ \bibnamefont {Qi}}, \ and\
  \bibinfo {author} {\bibfnamefont {D.}~\bibnamefont {Stanford}},\ }\href
  {\doibase 10.1007/JHEP05(2017)125} {\bibfield  {journal} {\bibinfo  {journal}
  {Journal of High Energy Physics}\ }\textbf {\bibinfo {volume} {05}},\
  \bibinfo {pages} {125} (\bibinfo {year} {2017})}\BibitemShut {NoStop}%
\bibitem [{\citenamefont {Davison}\ \emph {et~al.}(2017)\citenamefont
  {Davison}, \citenamefont {Fu}, \citenamefont {Georges}, \citenamefont {Gu},
  \citenamefont {Jensen},\ and\ \citenamefont {Sachdev}}]{davison2017}%
  \BibitemOpen
  \bibfield  {author} {\bibinfo {author} {\bibfnamefont {R.~A.}\ \bibnamefont
  {Davison}}, \bibinfo {author} {\bibfnamefont {W.}~\bibnamefont {Fu}},
  \bibinfo {author} {\bibfnamefont {A.}~\bibnamefont {Georges}}, \bibinfo
  {author} {\bibfnamefont {Y.}~\bibnamefont {Gu}}, \bibinfo {author}
  {\bibfnamefont {K.}~\bibnamefont {Jensen}}, \ and\ \bibinfo {author}
  {\bibfnamefont {S.}~\bibnamefont {Sachdev}},\ }\href {\doibase
  10.1103/PhysRevB.95.155131} {\bibfield  {journal} {\bibinfo  {journal} {Phys.
  Rev. B}\ }\textbf {\bibinfo {volume} {95}},\ \bibinfo {pages} {155131}
  (\bibinfo {year} {2017})}\BibitemShut {NoStop}%
\bibitem [{\citenamefont {Gross}\ and\ \citenamefont
  {Rosenhaus}(2017)}]{gross2017}%
  \BibitemOpen
  \bibfield  {author} {\bibinfo {author} {\bibfnamefont {D.~J.}\ \bibnamefont
  {Gross}}\ and\ \bibinfo {author} {\bibfnamefont {V.}~\bibnamefont
  {Rosenhaus}},\ }\href {\doibase 10.1007/JHEP02(2017)093} {\bibfield
  {journal} {\bibinfo  {journal} {Journal of High Energy Physics}\ }\textbf
  {\bibinfo {volume} {02}},\ \bibinfo {pages} {93} (\bibinfo {year}
  {2017})}\BibitemShut {NoStop}%
\bibitem [{\citenamefont {Gout{\'e}raux}\ and\ \citenamefont
  {Kiritsis}(2011)}]{gouteraux2011}%
  \BibitemOpen
  \bibfield  {author} {\bibinfo {author} {\bibfnamefont {B.}~\bibnamefont
  {Gout{\'e}raux}}\ and\ \bibinfo {author} {\bibfnamefont {E.}~\bibnamefont
  {Kiritsis}},\ }\href {\doibase 10.1007/JHEP12(2011)036} {\bibfield  {journal}
  {\bibinfo  {journal} {Journal of High Energy Physics}\ }\textbf {\bibinfo
  {volume} {12}},\ \bibinfo {pages} {36} (\bibinfo {year} {2011})}\BibitemShut
  {NoStop}%
\bibitem [{\citenamefont {Guhr}\ \emph {et~al.}(1998)\citenamefont {Guhr},
  \citenamefont {Mueller-Groeling},\ and\ \citenamefont
  {Weidenmueller}}]{guhr1998}%
  \BibitemOpen
  \bibfield  {author} {\bibinfo {author} {\bibfnamefont {T.}~\bibnamefont
  {Guhr}}, \bibinfo {author} {\bibfnamefont {A.}~\bibnamefont
  {Mueller-Groeling}}, \ and\ \bibinfo {author} {\bibfnamefont {H.~A.}\
  \bibnamefont {Weidenmueller}},\ }\href {\doibase
  http://dx.doi.org/10.1016/S0370-1573(97)00088-4} {\bibfield  {journal}
  {\bibinfo  {journal} {Physics Reports}\ }\textbf {\bibinfo {volume} {299}},\
  \bibinfo {pages} {189 } (\bibinfo {year} {1998})}\BibitemShut {NoStop}%
\bibitem [{\citenamefont {Alhassid}\ and\ \citenamefont
  {Levine}(1992)}]{alhassid1992}%
  \BibitemOpen
  \bibfield  {author} {\bibinfo {author} {\bibfnamefont {Y.}~\bibnamefont
  {Alhassid}}\ and\ \bibinfo {author} {\bibfnamefont {R.~D.}\ \bibnamefont
  {Levine}},\ }\href {\doibase 10.1103/PhysRevA.46.4650} {\bibfield  {journal}
  {\bibinfo  {journal} {Phys. Rev. A}\ }\textbf {\bibinfo {volume} {46}},\
  \bibinfo {pages} {4650} (\bibinfo {year} {1992})}\BibitemShut {NoStop}%
\bibitem [{\citenamefont {Torres-Herrera}\ \emph {et~al.}(2017)\citenamefont
  {Torres-Herrera}, \citenamefont {Garc{\'\i}a-Garc{\'\i}a},\ and\
  \citenamefont {Santos}}]{Torres-Herrera2017}%
  \BibitemOpen
  \bibfield  {author} {\bibinfo {author} {\bibfnamefont {E.~J.}\ \bibnamefont
  {Torres-Herrera}}, \bibinfo {author} {\bibfnamefont {A.~M.}\ \bibnamefont
  {Garc{\'\i}a-Garc{\'\i}a}}, \ and\ \bibinfo {author} {\bibfnamefont {L.~F.}\
  \bibnamefont {Santos}},\ }\href@noop {} {\  (\bibinfo {year} {2017})},\
  \Eprint {http://arxiv.org/abs/1704.06272} {arXiv:1704.06272} \BibitemShut
  {NoStop}%
\bibitem [{\citenamefont {Mehta}(2004)}]{mehta2004}%
  \BibitemOpen
  \bibfield  {author} {\bibinfo {author} {\bibfnamefont {M.~L.}\ \bibnamefont
  {Mehta}},\ }\href@noop {} {\emph {\bibinfo {title} {Random matrices}}}\
  (\bibinfo  {publisher} {Academic press},\ \bibinfo {year} {2004})\BibitemShut
  {NoStop}%
\bibitem [{\citenamefont {Luitz}\ \emph {et~al.}(2015)\citenamefont {Luitz},
  \citenamefont {Laflorencie},\ and\ \citenamefont {Alet}}]{luitz2015}%
  \BibitemOpen
  \bibfield  {author} {\bibinfo {author} {\bibfnamefont {D.~J.}\ \bibnamefont
  {Luitz}}, \bibinfo {author} {\bibfnamefont {N.}~\bibnamefont {Laflorencie}},
  \ and\ \bibinfo {author} {\bibfnamefont {F.}~\bibnamefont {Alet}},\ }\href
  {\doibase 10.1103/PhysRevB.91.081103} {\bibfield  {journal} {\bibinfo
  {journal} {Phys. Rev. B}\ }\textbf {\bibinfo {volume} {91}},\ \bibinfo
  {pages} {081103} (\bibinfo {year} {2015})}\BibitemShut {NoStop}%
\bibitem [{\citenamefont {Oganesyan}\ and\ \citenamefont
  {Huse}(2007)}]{oganesyan2007}%
  \BibitemOpen
  \bibfield  {author} {\bibinfo {author} {\bibfnamefont {V.}~\bibnamefont
  {Oganesyan}}\ and\ \bibinfo {author} {\bibfnamefont {D.~A.}\ \bibnamefont
  {Huse}},\ }\href {\doibase 10.1103/PhysRevB.75.155111} {\bibfield  {journal}
  {\bibinfo  {journal} {Phys. Rev. B}\ }\textbf {\bibinfo {volume} {75}},\
  \bibinfo {pages} {155111} (\bibinfo {year} {2007})}\BibitemShut {NoStop}%
\bibitem [{\citenamefont {Bertrand}\ and\ \citenamefont
  {Garc\'{\i}a-Garc\'{\i}a}(2016)}]{bertrand2016}%
  \BibitemOpen
  \bibfield  {author} {\bibinfo {author} {\bibfnamefont {C.~L.}\ \bibnamefont
  {Bertrand}}\ and\ \bibinfo {author} {\bibfnamefont {A.~M.}\ \bibnamefont
  {Garc\'{\i}a-Garc\'{\i}a}},\ }\href {\doibase 10.1103/PhysRevB.94.144201}
  {\bibfield  {journal} {\bibinfo  {journal} {Phys. Rev. B}\ }\textbf {\bibinfo
  {volume} {94}},\ \bibinfo {pages} {144201} (\bibinfo {year}
  {2016})}\BibitemShut {NoStop}%
\bibitem [{\citenamefont {Atas}\ \emph {et~al.}(2013)\citenamefont {Atas},
  \citenamefont {Bogomolny}, \citenamefont {Giraud},\ and\ \citenamefont
  {Roux}}]{atas2016}%
  \BibitemOpen
  \bibfield  {author} {\bibinfo {author} {\bibfnamefont {Y.~Y.}\ \bibnamefont
  {Atas}}, \bibinfo {author} {\bibfnamefont {E.}~\bibnamefont {Bogomolny}},
  \bibinfo {author} {\bibfnamefont {O.}~\bibnamefont {Giraud}}, \ and\ \bibinfo
  {author} {\bibfnamefont {G.}~\bibnamefont {Roux}},\ }\href {\doibase
  10.1103/PhysRevLett.110.084101} {\bibfield  {journal} {\bibinfo  {journal}
  {Phys. Rev. Lett.}\ }\textbf {\bibinfo {volume} {110}},\ \bibinfo {pages}
  {084101} (\bibinfo {year} {2013})}\BibitemShut {NoStop}%
\bibitem [{\citenamefont {Kudrolli}\ \emph {et~al.}(1994)\citenamefont
  {Kudrolli}, \citenamefont {Sridhar}, \citenamefont {Pandey},\ and\
  \citenamefont {Ramaswamy}}]{kudrolli1994}%
  \BibitemOpen
  \bibfield  {author} {\bibinfo {author} {\bibfnamefont {A.}~\bibnamefont
  {Kudrolli}}, \bibinfo {author} {\bibfnamefont {S.}~\bibnamefont {Sridhar}},
  \bibinfo {author} {\bibfnamefont {A.}~\bibnamefont {Pandey}}, \ and\ \bibinfo
  {author} {\bibfnamefont {R.}~\bibnamefont {Ramaswamy}},\ }\href {\doibase
  10.1103/PhysRevE.49.R11} {\bibfield  {journal} {\bibinfo  {journal} {Phys.
  Rev. E}\ }\textbf {\bibinfo {volume} {49}},\ \bibinfo {pages} {R11} (\bibinfo
  {year} {1994})}\BibitemShut {NoStop}%
\bibitem [{\citenamefont {Alt}\ \emph {et~al.}(1997)\citenamefont {Alt},
  \citenamefont {Gr\"af}, \citenamefont {Guhr}, \citenamefont {Harney},
  \citenamefont {Hofferbert}, \citenamefont {Rehfeld}, \citenamefont
  {Richter},\ and\ \citenamefont {Schardt}}]{alt1997}%
  \BibitemOpen
  \bibfield  {author} {\bibinfo {author} {\bibfnamefont {H.}~\bibnamefont
  {Alt}}, \bibinfo {author} {\bibfnamefont {H.-D.}\ \bibnamefont {Gr\"af}},
  \bibinfo {author} {\bibfnamefont {T.}~\bibnamefont {Guhr}}, \bibinfo {author}
  {\bibfnamefont {H.~L.}\ \bibnamefont {Harney}}, \bibinfo {author}
  {\bibfnamefont {R.}~\bibnamefont {Hofferbert}}, \bibinfo {author}
  {\bibfnamefont {H.}~\bibnamefont {Rehfeld}}, \bibinfo {author} {\bibfnamefont
  {A.}~\bibnamefont {Richter}}, \ and\ \bibinfo {author} {\bibfnamefont
  {P.}~\bibnamefont {Schardt}},\ }\href {\doibase 10.1103/PhysRevE.55.6674}
  {\bibfield  {journal} {\bibinfo  {journal} {Phys. Rev. E}\ }\textbf {\bibinfo
  {volume} {55}},\ \bibinfo {pages} {6674} (\bibinfo {year}
  {1997})}\BibitemShut {NoStop}%
\bibitem [{\citenamefont {\AA{}berg}(1990)}]{aberg1990}%
  \BibitemOpen
  \bibfield  {author} {\bibinfo {author} {\bibfnamefont {S.}~\bibnamefont
  {\AA{}berg}},\ }\href {\doibase 10.1103/PhysRevLett.64.3119} {\bibfield
  {journal} {\bibinfo  {journal} {Phys. Rev. Lett.}\ }\textbf {\bibinfo
  {volume} {64}},\ \bibinfo {pages} {3119} (\bibinfo {year}
  {1990})}\BibitemShut {NoStop}%
\bibitem [{\citenamefont {Jacquod}\ and\ \citenamefont
  {Shepelyansky}(1997)}]{jacquod1997}%
  \BibitemOpen
  \bibfield  {author} {\bibinfo {author} {\bibfnamefont {P.}~\bibnamefont
  {Jacquod}}\ and\ \bibinfo {author} {\bibfnamefont {D.~L.}\ \bibnamefont
  {Shepelyansky}},\ }\href {\doibase 10.1103/PhysRevLett.79.1837} {\bibfield
  {journal} {\bibinfo  {journal} {Phys. Rev. Lett.}\ }\textbf {\bibinfo
  {volume} {79}},\ \bibinfo {pages} {1837} (\bibinfo {year}
  {1997})}\BibitemShut {NoStop}%
\bibitem [{\citenamefont {Berkovits}\ and\ \citenamefont
  {Avishai}(1998)}]{berkovits1998}%
  \BibitemOpen
  \bibfield  {author} {\bibinfo {author} {\bibfnamefont {R.}~\bibnamefont
  {Berkovits}}\ and\ \bibinfo {author} {\bibfnamefont {Y.}~\bibnamefont
  {Avishai}},\ }\href {\doibase 10.1103/PhysRevLett.80.568} {\bibfield
  {journal} {\bibinfo  {journal} {Phys. Rev. Lett.}\ }\textbf {\bibinfo
  {volume} {80}},\ \bibinfo {pages} {568} (\bibinfo {year} {1998})}\BibitemShut
  {NoStop}%
\bibitem [{\citenamefont {Kota}\ \emph {et~al.}(2011)\citenamefont {Kota},
  \citenamefont {Rela{\~n}o}, \citenamefont {Retamosa},\ and\ \citenamefont
  {Vyas}}]{kota2011a}%
  \BibitemOpen
  \bibfield  {author} {\bibinfo {author} {\bibfnamefont {V.~K.~B.}\
  \bibnamefont {Kota}}, \bibinfo {author} {\bibfnamefont {A.}~\bibnamefont
  {Rela{\~n}o}}, \bibinfo {author} {\bibfnamefont {J.}~\bibnamefont
  {Retamosa}}, \ and\ \bibinfo {author} {\bibfnamefont {M.}~\bibnamefont
  {Vyas}},\ }\href {http://stacks.iop.org/1742-5468/2011/i=10/a=P10028}
  {\bibfield  {journal} {\bibinfo  {journal} {Journal of Statistical Mechanics:
  Theory and Experiment}\ }\textbf {\bibinfo {volume} {2011}},\ \bibinfo
  {pages} {P10028} (\bibinfo {year} {2011})}\BibitemShut {NoStop}%
\bibitem [{\citenamefont {Di~Stasio}\ and\ \citenamefont
  {Zotos}(1995)}]{stasio1995}%
  \BibitemOpen
  \bibfield  {author} {\bibinfo {author} {\bibfnamefont {M.}~\bibnamefont
  {Di~Stasio}}\ and\ \bibinfo {author} {\bibfnamefont {X.}~\bibnamefont
  {Zotos}},\ }\href {\doibase 10.1103/PhysRevLett.74.2050} {\bibfield
  {journal} {\bibinfo  {journal} {Phys. Rev. Lett.}\ }\textbf {\bibinfo
  {volume} {74}},\ \bibinfo {pages} {2050} (\bibinfo {year}
  {1995})}\BibitemShut {NoStop}%
\bibitem [{\citenamefont {Sekino}\ and\ \citenamefont
  {Susskind}(2008)}]{sekino2008}%
  \BibitemOpen
  \bibfield  {author} {\bibinfo {author} {\bibfnamefont {Y.}~\bibnamefont
  {Sekino}}\ and\ \bibinfo {author} {\bibfnamefont {L.}~\bibnamefont
  {Susskind}},\ }\href@noop {} {\bibfield  {journal} {\bibinfo  {journal}
  {Journal of High Energy Physics}\ }\textbf {\bibinfo {volume} {10}},\
  \bibinfo {pages} {065} (\bibinfo {year} {2008})}\BibitemShut {NoStop}%
\bibitem [{\citenamefont {Maldacena}\ \emph
  {et~al.}(2016{\natexlab{b}})\citenamefont {Maldacena}, \citenamefont
  {Shenker},\ and\ \citenamefont {Stanford}}]{maldacena2015}%
  \BibitemOpen
  \bibfield  {author} {\bibinfo {author} {\bibfnamefont {J.}~\bibnamefont
  {Maldacena}}, \bibinfo {author} {\bibfnamefont {S.~H.}\ \bibnamefont
  {Shenker}}, \ and\ \bibinfo {author} {\bibfnamefont {D.}~\bibnamefont
  {Stanford}},\ }\href {\doibase 10.1007/JHEP08(2016)106} {\bibfield  {journal}
  {\bibinfo  {journal} {Journal of High Energy Physics}\ }\textbf {\bibinfo
  {volume} {08}},\ \bibinfo {pages} {106} (\bibinfo {year}
  {2016}{\natexlab{b}})}\BibitemShut {NoStop}%
\bibitem [{\citenamefont {Larkin}\ and\ \citenamefont
  {Ovchinnikov}(1969)}]{larkin1969}%
  \BibitemOpen
  \bibfield  {author} {\bibinfo {author} {\bibfnamefont {A.}~\bibnamefont
  {Larkin}}\ and\ \bibinfo {author} {\bibfnamefont {Y.~N.}\ \bibnamefont
  {Ovchinnikov}},\ }\href@noop {} {\bibfield  {journal} {\bibinfo  {journal}
  {Sov Phys JETP}\ }\textbf {\bibinfo {volume} {28}},\ \bibinfo {pages} {1200}
  (\bibinfo {year} {1969})}\BibitemShut {NoStop}%
\bibitem [{\citenamefont {Berman}\ and\ \citenamefont
  {Zaslavsky}(1978)}]{berman1978}%
  \BibitemOpen
  \bibfield  {author} {\bibinfo {author} {\bibfnamefont {G.}~\bibnamefont
  {Berman}}\ and\ \bibinfo {author} {\bibfnamefont {G.}~\bibnamefont
  {Zaslavsky}},\ }\href {\doibase
  http://dx.doi.org/10.1016/0378-4371(78)90190-5} {\bibfield  {journal}
  {\bibinfo  {journal} {Physica A: Statistical Mechanics and its Applications}\
  }\textbf {\bibinfo {volume} {91}},\ \bibinfo {pages} {450 } (\bibinfo {year}
  {1978})}\BibitemShut {NoStop}%
\bibitem [{\citenamefont {Lai}\ \emph {et~al.}(1993)\citenamefont {Lai},
  \citenamefont {Ott},\ and\ \citenamefont {Grebogi}}]{lai1993}%
  \BibitemOpen
  \bibfield  {author} {\bibinfo {author} {\bibfnamefont {Y.}~\bibnamefont
  {Lai}}, \bibinfo {author} {\bibfnamefont {E.}~\bibnamefont {Ott}}, \ and\
  \bibinfo {author} {\bibfnamefont {C.}~\bibnamefont {Grebogi}},\ }\href
  {\doibase 10.1016/0375-9601(93)90178-3} {\bibfield  {journal} {\bibinfo
  {journal} {Physics Letters, Section A: General, Atomic and Solid State
  Physics}\ }\textbf {\bibinfo {volume} {173}},\ \bibinfo {pages} {148}
  (\bibinfo {year} {1993})}\BibitemShut {NoStop}%
\bibitem [{\citenamefont {Polchinski}\ and\ \citenamefont
  {Rosenhaus}(2016)}]{polchinski2016}%
  \BibitemOpen
  \bibfield  {author} {\bibinfo {author} {\bibfnamefont {J.}~\bibnamefont
  {Polchinski}}\ and\ \bibinfo {author} {\bibfnamefont {V.}~\bibnamefont
  {Rosenhaus}},\ }\href {\doibase 10.1007/JHEP04(2016)001} {\bibfield
  {journal} {\bibinfo  {journal} {Journal of High Energy Physics}\ }\textbf
  {\bibinfo {volume} {04}},\ \bibinfo {pages} {1} (\bibinfo {year}
  {2016})}\BibitemShut {NoStop}%
\bibitem [{\citenamefont {Song}\ \emph {et~al.}(2017)\citenamefont {Song},
  \citenamefont {Jian},\ and\ \citenamefont {Balents}}]{song2017}%
  \BibitemOpen
  \bibfield  {author} {\bibinfo {author} {\bibfnamefont {X.-Y.}\ \bibnamefont
  {Song}}, \bibinfo {author} {\bibfnamefont {C.-M.}\ \bibnamefont {Jian}}, \
  and\ \bibinfo {author} {\bibfnamefont {L.}~\bibnamefont {Balents}},\ }\href
  {\doibase 10.1103/PhysRevLett.119.216601} {\bibfield  {journal} {\bibinfo
  {journal} {Phys. Rev. Lett.}\ }\textbf {\bibinfo {volume} {119}},\ \bibinfo
  {pages} {216601} (\bibinfo {year} {2017})}\BibitemShut {NoStop}%
\bibitem [{\citenamefont {Chen}\ \emph {et~al.}(2017)\citenamefont {Chen},
  \citenamefont {Fan}, \citenamefont {Chen}, \citenamefont {Zhai},\ and\
  \citenamefont {Zhang}}]{chen2017}%
  \BibitemOpen
  \bibfield  {author} {\bibinfo {author} {\bibfnamefont {X.}~\bibnamefont
  {Chen}}, \bibinfo {author} {\bibfnamefont {R.}~\bibnamefont {Fan}}, \bibinfo
  {author} {\bibfnamefont {Y.}~\bibnamefont {Chen}}, \bibinfo {author}
  {\bibfnamefont {H.}~\bibnamefont {Zhai}}, \ and\ \bibinfo {author}
  {\bibfnamefont {P.}~\bibnamefont {Zhang}},\ }\href {\doibase
  10.1103/PhysRevLett.119.207603} {\bibfield  {journal} {\bibinfo  {journal}
  {Phys. Rev. Lett.}\ }\textbf {\bibinfo {volume} {119}},\ \bibinfo {pages}
  {207603} (\bibinfo {year} {2017})}\BibitemShut {NoStop}%
\bibitem [{\citenamefont {Eberlein}\ \emph {et~al.}(2017)\citenamefont
  {Eberlein}, \citenamefont {Kasper}, \citenamefont {Sachdev},\ and\
  \citenamefont {Steinberg}}]{eberlein2017}%
  \BibitemOpen
  \bibfield  {author} {\bibinfo {author} {\bibfnamefont {A.}~\bibnamefont
  {Eberlein}}, \bibinfo {author} {\bibfnamefont {V.}~\bibnamefont {Kasper}},
  \bibinfo {author} {\bibfnamefont {S.}~\bibnamefont {Sachdev}}, \ and\
  \bibinfo {author} {\bibfnamefont {J.}~\bibnamefont {Steinberg}},\ }\href
  {\doibase 10.1103/PhysRevB.96.205123} {\bibfield  {journal} {\bibinfo
  {journal} {Phys. Rev. B}\ }\textbf {\bibinfo {volume} {96}},\ \bibinfo
  {pages} {205123} (\bibinfo {year} {2017})}\BibitemShut {NoStop}%
\bibitem [{Note1()}]{Note1}%
  \BibitemOpen
  \bibinfo {note} {The reason is simply that for finite $N$ there is a gap of
  the order $2^{-N}$ so that degeneracy is not exact.}\BibitemShut {Stop}%
\bibitem [{Note2()}]{Note2}%
  \BibitemOpen
  \bibinfo {note} {Denoting the Hamiltonian in Eq.\protect \textup {\hbox
  {\mathsurround \z@ \protect \normalfont (\ignorespaces \ref {hami}\unskip
  \@@italiccorr )}} by $H(\kappa )$, we have that $\protect \mathaccentV
  {tilde}07E{H}(\protect \mathaccentV {tilde}07E{\kappa }) = \protect
  \mathaccentV {tilde}07E{\kappa }^{-1}H(\protect \mathaccentV
  {tilde}07E{\kappa }^{-1})$ with $J=1$.}\BibitemShut {Stop}%
\bibitem [{Note3()}]{Note3}%
  \BibitemOpen
  \bibinfo {note} {Note that since $\rho $ is even, without loss of generality
  we can take a symmetric interval $(-\lambda ,\lambda )$.}\BibitemShut {Stop}%
\bibitem [{Note4()}]{Note4}%
  \BibitemOpen
  \bibinfo {note} {We have also checked numerically that, as long and
  positivity of $\rho $ is respected, changes in the cutoff do not affect the
  results for the thermodynamic coefficients $c$ and $s_0$.}\BibitemShut
  {Stop}%
\end{thebibliography}%

\end{document}